\documentstyle[12pt,aasms]{article}
\font\smfont=cmr7
\font\smtfont=cmr5
\font\smttfont=cmr5

\newfam\smfam

\textfont\smfam=\smfont
\scriptfont\smfam=\smtfont
\scriptscriptfont\smfam=\smttfont

\def\spose#1{\hbox to 0pt{#1\hss}}

\def\kms{\ifmmode {\rm\,km\,s^{-1}}\else
    ${\rm\,km\,s^{-1}}$\fi}
\def\kmsMpc{\ifmmode {\rm\,km\,s^{-1}\,Mpc^{-1}}\else
    ${\rm\,km\,s^{-1}\,Mpc^{-1}}$\fi}

\def\msun{\ifmmode {\rm\,M_\odot}\else ${\rm\,M_\odot}$\fi}
\def\Msun{\ifmmode {\rm\,M_\odot}\else ${\rm\,M_\odot}$\fi}
\def\lsun{\ifmmode {\rm\,L_\odot}\else ${\rm\,L_\odot}$\fi}
\def\Lsun{\ifmmode {\rm\,L_\odot}\else ${\rm\,L_\odot}$\fi}
\def\rsun{\ifmmode {\rm\,R_\odot}\else ${\rm\,R_\odot}$\fi}
\def\Rsun{\ifmmode {\rm\,R_\odot}\else ${\rm\,R_\odot}$\fi}

\def\cm{{\rm\,cm}}
\def\cm3{\ifmmode {\rm\,cm^{-3}}\else ${\rm\,cm^{-3}}$\fi}

\def\ergps{\ifmmode {\rm\,erg\,s^{-1}}\else ${\rm\,erg\,s^{-1}}$\fi}
\def\ergpscm2{\ifmmode {\rm\,erg\,s^{-1}\,cm^{-2}}\else
    ${\rm\,erg\,s^{-1}\,cm^{-2}}$\fi}

\def\eg{{\it e.g.}}
\def\deg{\ifmmode {^{\circ}}\else {$^\circ$}\fi}
\def\degr{\ifmmode {^{\circ}}\else {$^\circ$}\fi}
\def\degs{\ifmmode {^{\circ}}\else {$^\circ$}\fi}

\def\etal{{\it et al.~}}

\def\h3Mpc{h^{-3}{\rm Mpc}^3}
\def\Ho{\ifmmode {\rm\,H_\circ}\else ${\rm\,H_\circ}$\fi}
\def\hnot{\ifmmode {\rm\,H_\circ}\else ${\rm\,H_\circ}$\fi}
\def\h0{\ifmmode {\rm\,H_\circ}\else ${\rm\,H_\circ}$\fi}
\def\hnotunit{\ifmmode {\rm\,km\,s^{-1}\,Mpc^{-1}}\else
    ${\rm\,km\,s^{-1}\,Mpc^{-1}}$\fi}
\def\qnot{\ifmmode {\rm\,q_\circ}\else ${\rm q_\circ}$\fi}
\def\q0{\ifmmode {\rm\,q_\circ}\else ${\rm q_\circ}$\fi}
\def\ie{{\it i.e.}}

\def\arcsec{\ifmmode {^{\prime\prime}}\else $^{\prime\prime}$\fi}
\def\asec{\ifmmode {^{\prime\prime}}\else $^{\prime\prime}$\fi}
\def\arcmin{\ifmmode {^{\prime}}\else $^{\prime}$\fi}
\def\amin{\ifmmode {^{\prime}}\else $^{\prime}$\fi}

\def\secper{\ifmmode \rlap.{^{s}}\else $\rlap{.}{^{s}} $\fi}
\def\minper{\ifmmode \rlap.{^{m}}\else $\rlap{.}{^m} $\fi}
\def\magper{\ifmmode \rlap.{^{m}}\else $\rlap{.}{^m} $\fi}
\def\arcsper{\ifmmode \rlap.{^{\prime\prime}}\else
    $\rlap.{^{\prime\prime}}$\fi}
\def\arcmper{\ifmmode \rlap.{^{\prime}}\else
    $\rlap.{^{\prime}}$\fi}
\def\spose#1{\hbox to 0pt{#1\hss}}
\def\simlt{\mathrel{\spose{\lower 3pt\hbox{$\mathchar"218$}}
     \raise 2.0pt\hbox{$\mathchar"13C$}}}
\def\simgt{\mathrel{\spose{\lower 3pt\hbox{$\mathchar"218$}}
     \raise 2.0pt\hbox{$\mathchar"13E$}}}

\def\refindent{\par\noindent\parskip=2pt\hangindent=3pc\hangafter=1 }

\def\araa{{ARA\&A}}
\def\aa{{A\&A}}

\def\aj{{AJ}}
\def\apj{{ApJ}}
\def\apjlett{{ApJ}}
\def\apjl{{ApJ}}
\def\apjsupp{{ApJS}}
\def\apjs{{ApJS}}

\def\mn{{MNRAS}}
\def\mnras{{MNRAS}}
\def\nature{{Nature}}

\def\pasp{{PASP}}

\def\apjref#1;#2;#3;#4 {\par\pp#1, {#2}, #3, #4 \par}

\tighten

\received{09 August 1995}
\accepted{ 11 January 1996}
\journalid{465}{1 July 1996}

\begin{document}

\title{On the Origin of the UV Continuum Emission from 
the High Redshift Radio Galaxy 3C256\altaffilmark{1}}
\author{Arjun Dey\altaffilmark{2}}
\affil{Astronomy Dept., University of California at Berkeley, CA 94720, \&}
\affil{Institute of Geophysics \& Planetary Physics, LLNL, Livermore, CA 94550}
\affil{dey@noao.edu}
\author{Andrea Cimatti\altaffilmark{3} \& Wil van Breugel}
\affil{Institute of Geophysics \& Planetary Physics, LLNL, Livermore, CA 94550}
\affil{cimatti/wil@igpp.llnl.gov}
\author{Robert Antonucci}
\affil{Physics Dept., University of California, Santa Barbara, CA 93106}
\affil{ski@chester.physics.ucsb.edu}
\author{Hyron Spinrad}
\affil{Astronomy Dept., University of California at Berkeley, CA 94720}
\affil{spinrad@astro.berkeley.edu}
\medskip
\bigskip
\bigskip
\centerline{\it To appear in ApJ, July 1, 1996, vol. 465}
\bigskip
\bigskip
\altaffiltext{1}{Based on observations at the W.\ M.\ Keck Observatory.}
\altaffiltext{2}{Present address: NOAO/KPNO, 950 N.\ Cherry Ave., P.O.\ Box
26732, Tucson, AZ 85726; E-mail: dey@noao.edu}
\altaffiltext{3}{On leave from Osservatorio Astrofisico di Arcetri, Largo E.\ 
Fermi 5, I--50125, Firenze, Italy; E-mail: cimatti@arcetri.astro.it}

\begin{abstract}

We report spectropolarimetric observations obtained with the
W.\ M.\ Keck Telescope of the high redshift (z=1.824) aligned radio
galaxy 3C256. Our observations confirm that the spatially extended UV
continuum emission from this galaxy is polarized ($P_V\approx
10.9\%\pm0.9\%$) with the electric vector perpendicular to the aligned
radio and optical major axes ($\theta\approx 48.0^\circ\pm
2.4^\circ$).  This strongly suggests that a significant fraction of the
rest frame UV continuum emission from the galaxy is not starlight, but 
is instead scattered light
from a powerful AGN which is hidden from our direct view. The narrow
emission lines, including MgII, are unpolarized. The percentage
polarization of the continuum emission and the polarization position
angle are roughly constant as a function of wavelength. Although the 
present data do not permit us to discriminate between cool electrons 
and dust as the origin of the scattering, scattering 
by a population of hot ($T\simgt 10^7$~K) electrons cannot be the
dominant process, since such a population would overproduce X-ray
emission. A large population of cooler electrons ($T\approx 10^4$~K)
could be responsible for both the line emission and the scattered
light, but would require that the dust to gas ratio in the scattering
cones is $10^{-3}$ times smaller than that in our Galaxy and would
imply that a large fraction of the baryonic mass in the galaxy is in
the ionized component of its interstellar medium.  Dust scattering is
more efficient, but would result in detectable extinction of the emission
line spectrum unless the dust distribution is more highly clumped than
the line emitting gas.  Finally, we detect a strong
($W_\lambda^{rest}\approx 12$\AA) broad (FWHM $\approx$ 6500\kms)
absorption line centered at $\lambda_{rest} \approx 1477$\AA. We
discuss several possibilities for its origin and conclude that the most
likely candidate is absorption by a high velocity BAL cloud near the
nucleus of 3C256.

\end{abstract}


\section {Introduction}

Radio galaxies are often regarded as important probes of galaxy
evolution primarily because they are the only spatially extended,
`galaxy--like' objects that are detectable to very large lookback times.
Until recently, most evolutionary studies of these objects were founded
on the assumption that their observed optical continuum emission is
dominated by unreddened starlight (Bruzual 1986; Djorgovski, Spinrad, \&
Marr 1985). The uniformity of the observed $K-z$ Hubble diagram (\eg,
McCarthy 1993, Eales \etal 1993), and the fairly symmetric and
undisturbed near infrared morphologies of high redshift radio galaxies
seem to support this basic picture of radio galaxies as quiescently
evolving elliptical galaxies.

However, three discoveries made during the last decade have thrown this
complacent view of high redshift radio galaxies as young ellipticals
into considerable disarray. The first is that the observed optical
(rest frame UV) morphologies of high redshift radio galaxies are
aligned with their radio axes (McCarthy \etal 1987, Chambers \etal
1987), suggesting a causal relationship between the active nucleus and
the extended optical continuum emitting material ({\it e.g.,} Dunlop
and Peacock 1993). The second is the discovery that the rest frame UV
continuum emission is generally highly polarized with the electric
vector oriented perpendicular to the radio axis (di Serego Alighieri
\etal 1989, 1993, 1994; Jannuzi \& Elston 1991), which implies that a
significant fraction of the observed optical continuum emission is
nonstellar, and arises from processes related to the active nucleus.
In fact, di Serego Alighieri \etal (1994) detect a polarized, possibly broad 
MgII emission line, suggesting that the intrinsic nuclear source is 
indeed similar to that in quasars. 
Finally, the recent HST observations of the spectacular, complex
morphologies of high redshift radio galaxies (Miley \etal 1992,
Dickinson, Dey, \& Spinrad 1996, Dey \etal 1996a, Longair, Best, \&
R\"ottgering 1995) are in stark contradiction to the idea that these
objects are normal ellipticals. Although these data bode ill for
evolutionary studies of stellar populations in high redshift radio galaxies at
optical wavelengths, they provide the opportunity to investigate the
nature of the hidden active nucleus in these galaxies, the interactions
between the AGN and the ambient environment, and the relevance of these
interactions to the evolution of powerful AGN. They may also provide
constraints on the chemical evolution (and dust formation) timescale of
early galaxy evolution.

One question crucial to the study of these high redshift galaxies is:
what is the origin of their peculiar, aligned morphologies?
Several mechanisms have been proposed to explain the alignments (see
McCarthy 1993 for a review). Star formation triggered by the radio
source (DeYoung 1981, 1989, Rees 1990), scattering of light from a
hidden AGN off dust or electrons (di Serego Alighieri \etal 1989,
Fabian 1989), and inverse Compton scattering of cosmic microwave
background photons by relativistic electrons in or near the radio lobes (Daly
1992) appear to be the most relevant physical processes at high
redshift.  Spectropolarimetric observations can help distinguish
between these processes:  if the continuum emission is dominated by
unreddened starlight from a young population of stars, the continuum should be
unpolarized and should show stellar absorption features.
Alternatively, if scattered AGN light is the dominant constituent of
the extended continuum light, then the continuum should be highly
polarized with the electric vector perpendicular to the optical major
axis, and the spectrum of the polarized flux should be similar to that
of a quasar showing broad emission lines.  In addition, if the
mechanism responsible for the alignment effect is scattering,
spectropolarimetric observations of high redshift galaxies provide us
with the ability to investigate the properties of gas and dust in the 
environs of these high redshift galaxies at a large lookback
time.  For inverse Compton scattering, the continuum polarization
is likely to be low, and there should be a fairly close spatial
coincidence between the optical/UV rest frame continuum and the low
frequency radio continuum emission, which is generally not observed
({\it e.g.,} Miley \etal 1992, Dickinson, Dey, \& Spinrad 1995, Dey
\etal 1996a, Longair, Best, \& R\"ottgering 1995).

In this paper, we present 
spectropolarimetric observations of the high redshift (z=1.824) radio
galaxy 3C256 obtained using the W.\ M.\ Keck Telescope.  This paper is
the first missive from an ongoing project to determine the origin of
the UV continua of high redshift radio galaxies, in an effort to
understand their morphological and evolutionary properties.

3C256 is a particularly good
target for a detailed spectropolarimetric study for several reasons. 
First,
3C256 is an excellent example of the alignment effect: the optical
(rest frame UV) morphology is well aligned with the radio axis, has
a relatively high surface brightness, and is spatially
extended over similar scales as the radio source (Fig.~1). 
Second, 
broad band imaging polarimetric observations have shown that the
continuum emission from 3C256 is polarized and spatially
extended (Jannuzi 1994, Jannuzi \etal 1995).  3C256 can therefore
provide an important test of the AGN unification hypothesis, which
attributes the different observed properties of radio galaxies and
quasars primarily to the effects of orientation and obscuration (see
Antonucci 1993, and references therein).  If the continuum emission is
dominated by scattered light from a hidden quasar nucleus, then
spectropolarimetric observations allow a search for the broad line
component in the scattered flux.
Third, 
3C256 is considered to be a candidate protogalaxy because of its large
Ly$\alpha$ luminosity, a blue continuum spectral energy
distribution, similar optical and infrared morphologies (Spinrad \& 
Djorgovski 1984, Spinrad \etal 1985, Eisenhardt, personal communication),
and a $K$ magnitude that is $\sim$ 1 mag fainter than that predicted by
the mean $K-z$ relation for radio galaxies (Eisenhardt \& Dickinson 1992, 
Eales \etal 1993).
Since the integrated continuum from the galaxy is known to be polarized
(Jannuzi 1994, Jannuzi \etal 1995), it is important to determine the
true spectral energy distribution of the stellar population after
subtracting out the scattered component.
Finally, 
in a study of the coadded spectrum of high redshift radio galaxies, 
in which the signal was dominated by 3C256,
Chambers \& McCarthy (1990) reported the tentative detection of
absorption lines typical of young stellar populations and argued that
the rest frame UV continuum emission from the extended aligned component
in these objects is dominated by starlight. 
The deep total light spectrum that results from our co-added 
spectropolarimetric observations allows the most sensitive search to date for 
stellar absorption features in a single $z\approx 2$ radio galaxy. 

We describe our observations 
in section 2, and present our primary results in section 3. A discussion of 
these results and some preliminary spectroscopic information derived 
from the same data are presented in section 4. 

Throughout this paper we assume that \hnot=50\hnotunit\ and \qnot=0.1. 
The scale at $z=1.824$ (the redshift of 3C256) is then 11.3~kpc/\arcsec\ and 
the lookback time is nearly 12~Gyr, or 72\% of the age of the universe. 
In comparison, for \hnot=75\hnotunit\ and \qnot=0.5 the angular scale is 
5.6~kpc/\arcsec\ and the lookback time is 6.9~Gyr, or 79\% of the age of the 
universe. 

\section {Observations}

3C256 was observed using the Keck spectropolarimeter with the Low
Resolution Imaging Spectrometer (LRIS; Oke \etal 1995) at the
Cassegrain focus of the 10-m W. M. Keck Telescope on U.T. 1995 March 1.
The spectropolarimeter is a dual beam instrument which uses a calcite
analyser and a rotatable waveplate (see Goodrich \etal 1995 for
details). We used a 300 line/mm grating (blazed at $\lambda =
5000$\AA) to sample a wavelength range $\lambda\lambda$4000--9000\AA,
and a 1\arcsec\ wide slit which resulted in an effective resolution
FWHM of 10\AA.  The LRIS detector is a Tek 2048$^2$ CCD with 24$\mu$m
pixels which corresponds to a pixel scale of 0{\farcs}214 pix$^{-1}$.
The read noise was roughly 8.0 $e^-$ and the gain was $\approx 1.59
e^-$/adu.

We obtained three sets of observations with the spectrograph slit
oriented along the major axis of the optical continuum emission from
the galaxy (PA=140$^\circ$). Each set is comprised of observations made
in four waveplate positions ($0^\circ$, $45^\circ$, $22.5^\circ$,
$67.5^\circ$) which sample the electric vector in four position angles
on the sky ($140^\circ$, $50^\circ$, $5^\circ$, $95^\circ$
respectively). The exposure time per waveplate position was 1200$s$ for
the first two sets, and 900$s$ for the third set, resulting in a total
exposure time of 55 minutes per waveplate position. After each set, we 
reacquired the galaxy by using an offset from a nearby star. 
The seeing during our
observation was mediocre (1\arcsec\ to 1\farcs3), conditions were
non-photometric, and the maximum airmass was 1.55.

Since we observed the galaxy for about 4 hours, the parallactic angle
rotated during our observations from PA$_{parallactic} \approx
97$\deg\ to 180\deg (airmass 1.55 to 1.00).  Nevertheless, we estimate
that errors in relative spectrophotometry due to any differences
between the slit PA and the parallactic angle are small, since the
relative broad band photometry synthesised from our spectroscopy agrees
well with spectral energy distribution measured from broad band imaging
in $B$, $V$ and $R$ by Peter Eisenhardt and Richard Elston (Eisenhardt,
personal communication). The continuum in our spectral bandpass
($\lambda_{obs} 4000-9000$\AA) may be represented by a power
law $F_\nu\propto \nu^{-1.1\pm 0.1}$. The colors of the galaxy in a
1{\farcs}0$\times$3{\farcs}6 aperture derived from our spectroscopic
observations are $B-V\approx 0.3$, $V-R\approx0.5$, and $R-I\approx
0.7$. The $B$ and $I$ filters partly sample wavelength regions not
covered by our spectra, and therefore these magnitudes were derived by
smoothly extrapolating the observed spectrum to shorter and longer
wavelengths.  These colours are in fairly good agreement with the
Eisenhardt photometry and with the published $R-I=0.66\pm0.20$ color
measured by Le F\`evre \etal (1988).

In order to calibrate the instrumental polarization, we observed the
star G191B2B through UV and IR polaroid filters in four waveplate
positions. G191B2B is also a zero polarization standard (Schmidt,
Elston, \& Lupie 1992) as well as a flux calibration standard (see
below) and as an additional check on any residual polarization effects
we measured this star in the four waveplate positions. In order to
calibrate the polarization position angle zero point we observed the
polarization standard stars HD245310 and HD155528 (Schmidt, Elston, \&
Lupie 1992).  The Galactic latitude of 3C256 is 69\deg\ and the
interstellar percentage polarization is estimated to be very low
($<$0.2\%; Appenzeller 1968, Mathewson \& Ford 1970).

The data were corrected for overscan bias and flat--fielded using
internal lamps taken immediately following the observations.  The A--
and B--band (O$_2$) telluric absorption features were removed by
determining a template optical depth curve from the spectrum of a
standard star, scaling by the appropriate airmass, and dividing it into
the data.  The flux calibration was performed using observations of
G191B2B (Massey \etal 1988, Massey \& Gronwall 1990).  The star was
observed both with and without an order sorting GG495 filter in order
to correct for the second order light contamination in the spectral
region $\lambda > 7500$\AA.  All the spectroscopic reductions were
performed using the NOAO IRAF package. The spectropolarimetric analysis
was carried out using our own software, and is based on the methods
described in Miller, Robinson, \& Goodrich (1988).
The spectral modelling discussed in \S 4.2 was carried out 
using the SPECFIT package in IRAF (Kriss 1994).

Radio observations of 3C256 at 4.89~GHz and 14.96~GHz were obtained
using the VLA in A-array configuration on U.T.\ 17 August 1987.  The
4.89~GHz map is shown in figure \ref{radio}.  

\section {Results}

\subsection {Morphology of the Continuum and Line Emission}

\begin{figure}
\medskip
\plotfiddle{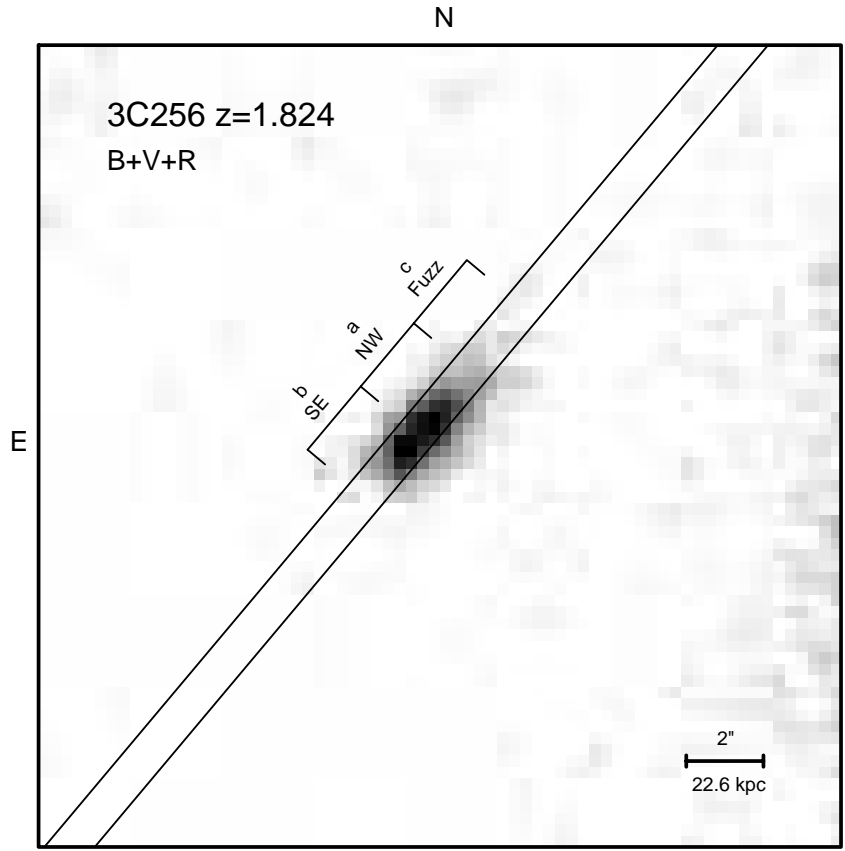}{7.0in}{0}{70}{70}{-230}{-20}
\caption{Broad band image ($B$+$V$+$R$) of 3C256 obtained by
S.\ Djorgovski \& H.\ Spinrad. The field of view is $\approx
21\arcsec\times 21\arcsec$. The optical galaxy position is
$\alpha_{1950}=11^h18^m04{\secper}2,\ \delta_{1950} =23^\circ 44^\prime
20{\farcs}5$.  The parallel lines denote the position and orientation
of the 1\arcsec\ slit used in our Keck LRIS spectropolarimetric
observations oriented in PA=140\deg. The bar to the NE of the galaxy
shows the approximate spectral extraction regions. The nomenclature of the 
`a', `b' and `c' components is the same as in Le F\`evre \etal 1988.}
\label{3c256bvr} 
\end{figure}

\begin{figure}
\plotone{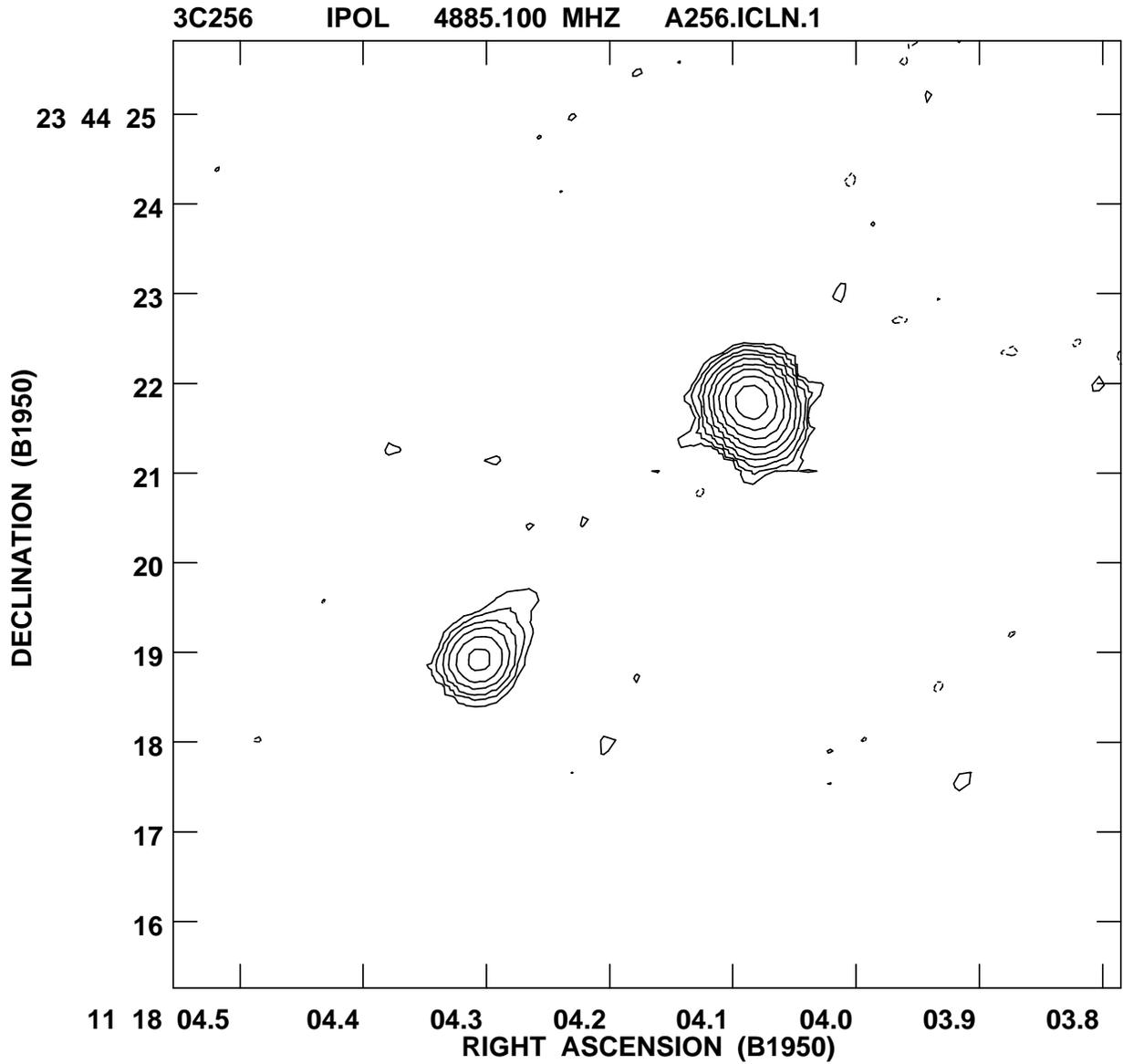}
\caption{VLA 4.885~GHz map of 3C~256. The
field of view is $10{\farcs}5\times 10{\farcs}5$. The contours are
plotted at flux densities of ($-$3,3,6,12,24,48,96,192,384,768)$\times\sigma$, 
where $\sigma=0.2$~mJy/beam.  The radio source is oriented at PA=140\deg. No 
radio core was detected at this sensitivity.}
\label{radio}
\end{figure}

As is seen in figure \ref{3c256bvr} (and more clearly in the higher
spatial resolution imaging observations presented by Le F\`evre \etal
1988), 3C256 is comprised of roughly three regions of optical emission
which are well aligned with the radio axis. The optical image shows two
clumps of emission in the south (`a' and `b' in Le F\`evre \etal 1988)
and a fainter, more diffuse component extending to the north west in
PA=140\deg.  The radio source has a double--lobed morphology
(fig.~\ref{radio}).  The radio lobes are separated by
$\approx$4\arcsec\ and the lobe flux ratio (north:south) is nearly 10:1
(Table 1).  No radio core is detected to a flux density
limit of 1.0~mJy (5$\sigma$ at 4.89~GHz).  Table~1 lists
the observed flux densities in the lobes.  The total flux measured from
the 4.89~GHz map agrees well with the single dish measurements ($F_{\rm
4.89~GHz}\approx 385\pm 48$~mJy; Gregory \& Condon 1991, Becker, White
\& Edwards 1991).  The radio lobes have extremely steep spectral 
indices, but this may be due to some fraction of the radio flux being 
resolved out of the 15~GHz map.

For comparison with the positions listed in Table 1, the optical
position of the galaxy is approximately $\alpha_{1950}\approx 11^h 18^m
04^s.2,\ \delta_{1950}\approx +23^\circ 44^\prime 21\farcs{7}$.
Although the relative astrometry between the radio and optical frames
is uncertain ($\pm 0{\farcs}8$), the morphological and angular size
differences between the radio and optical images ensure that there
cannot be a one-to-one correspondence between {\it both} radio lobes
and the regions of bright optical emission.  However, based on the
approximately similar size scales of the optical and radio extents, we
speculate that the fainter south eastern radio lobe may coincide with
the south eastern-most compact optical continuum emitting region
(component `b' in Le F\`evre \etal 1988) and the (much brighter)
northern radio lobe is associated with the faint, diffuse, extended
north western optical component (labelled `Fuzz' in
figure~\ref{3c256bvr}; component `c' in Le F\`evre \etal 1988).

\begin{figure}
\plotone{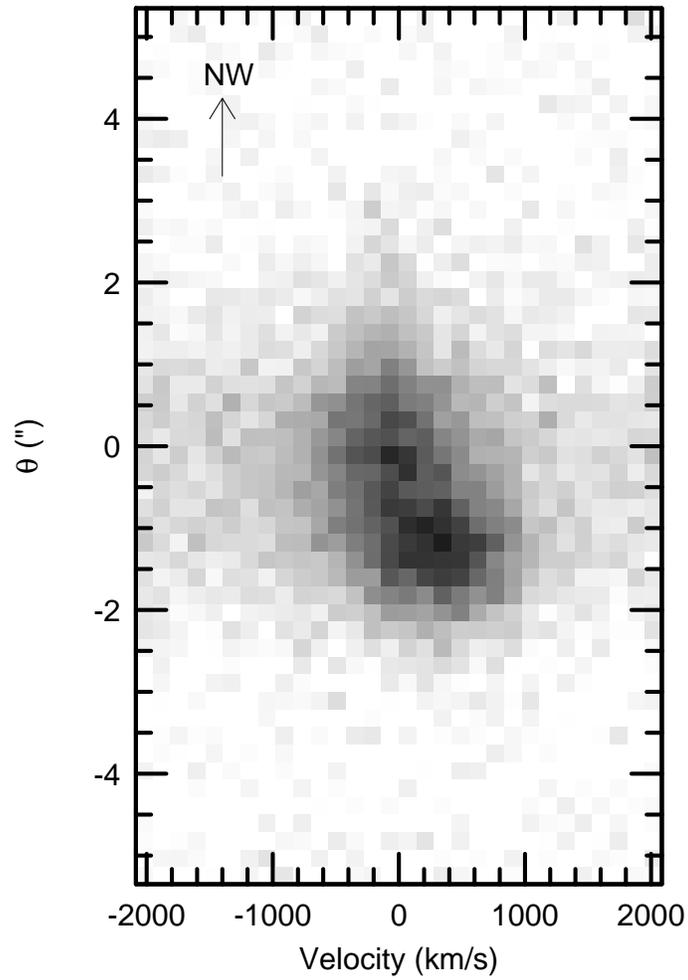}
\caption{Two dimensional spectrum of the CIII{]}$\lambda$1909 emission 
line in 3C256 obtained with the slit oriented at PA=140\deg.} 
\label{ciii}
\end{figure}

From our longslit spectra we find that 
the line emission shows a morphology similar to that seen in the Le
F\`evre \etal\ $R$ image: there are two clumps of emission in the south
east and a more diffuse faint component extending to the north west (figure
\ref{ciii}).
The brightest region of line emission is associated with the south
eastern region, and the faint, diffuse extended region of line emission
appears to be associated with the north western region. In this
respect, the emission line properties of 3C256 differ from those of
most other high redshift radio galaxies, where the brighter radio lobe
is usually associated with regions of bright optical emission
(McCarthy, van Breugel, \& Kapahi 1991).  

The velocity structure of the
emission line gas consists of three components.  The emission lines
associated with the south eastern component have broad line widths
(FWHM$\approx$1160\kms) whereas the central and north western regions
have relatively narrow line widths (FWHM$\approx$700\kms\ and 200\kms\ 
respectively) and show less velocity structure.  Overall, there is a
roughly monotonic velocity gradient along the major axis of the galaxy
with an end-to-end velocity difference of $\approx$400\kms. Note that 
the emission line region with the broadest line width (the SE region) 
is associated with the {\it fainter} radio lobe. Similar results are found 
in 3C324 and 3C368 (Dickinson \etal 1995, 
Dey \etal 1996a), where regions 
with the broadest line emission (FWHM$\sim$1000--1500\kms) are 
associated with the faintest radio components. 
\pagebreak

\subsection {One--Dimensional Imaging Polarimetry}

As in most high redshift radio galaxies, 
there is no clearly identifiable nucleus in either the optical or the
radio images of 3C256. Nevertheless, it is important to determine
whether the polarization in the galaxy arises from some unresolved
region, or whether the polarized flux is spatially extended. To address 
this question, we extracted cross-cuts of 3C256 by coadding our
spectropolarimetric data in the wavelength range
$\Delta\lambda =4000-6500$\AA\ ($\approx\lambda\lambda$1400--2300\AA\ 
in the rest frame).
The wavelength range was chosen to cover a large region of galaxian
emission where the signal-to-noise ratio was not adversely affected by
sky subtraction or fringing in the CCD.  These cross-cuts allow us to
determine the one dimensional spatial distribution (along the slit
PA=140\deg) of the percentage polarization and polarized flux.

\begin{figure} 
\plotfiddle{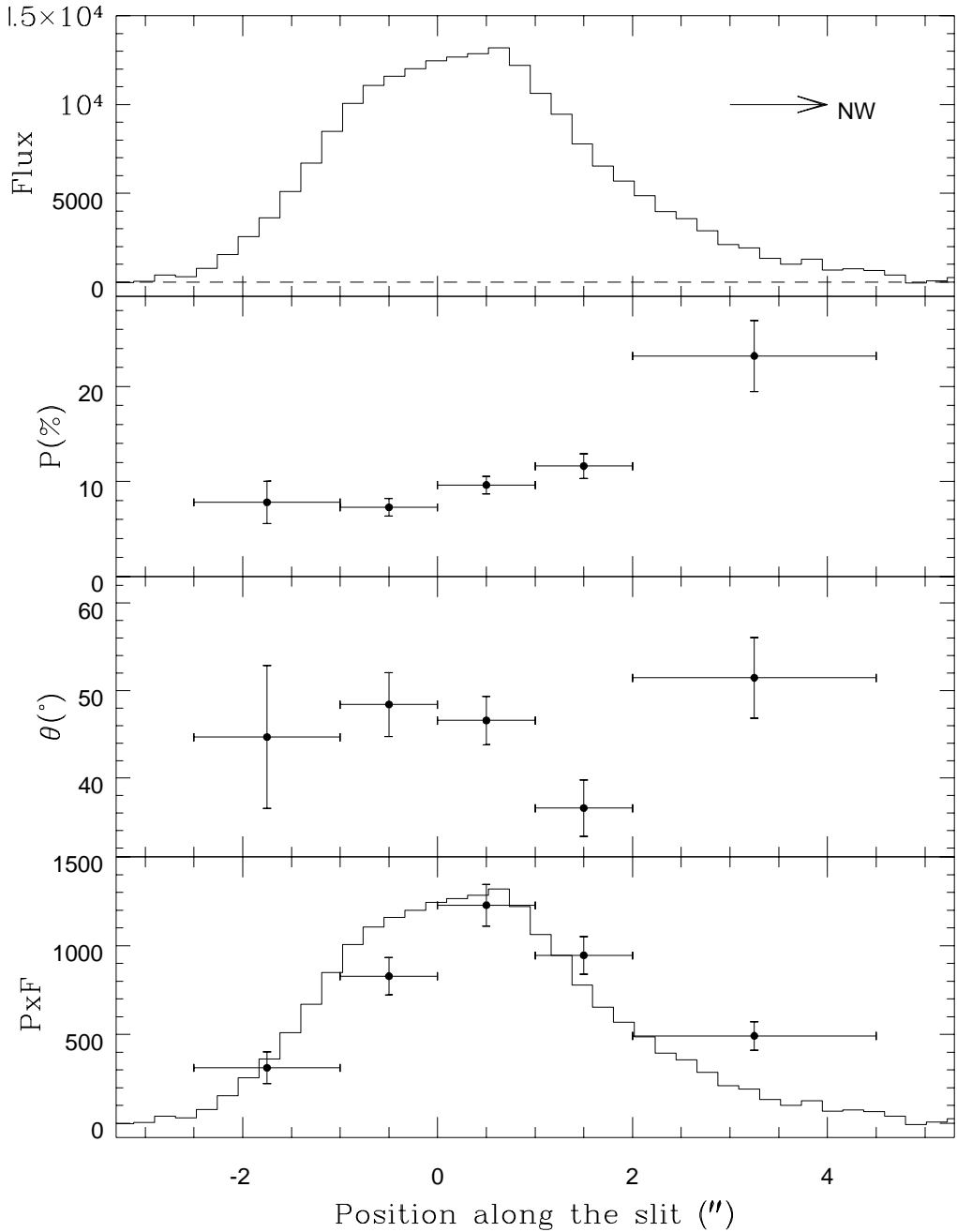}{7.0in}{0}{70}{70}{-230}{-20}
\vspace{0.1in}
\caption{The spatial variation along the slit (PA=140\deg) of the total
flux $F$ (in arbitrary units), 
percentage polarization $P$, polarization position angle
$\theta$, and polarized flux $P\times F$ for 3C256 measured in the
wavelength region $\lambda\lambda$4000--6500\AA\ ($\approx$1400--2300
in the rest frame). The abscissa is the position along the
slit in arcseconds relative to the center of the aperture used for the
wide spectroscopic extraction. $P$, $\theta$ and $P\times F$ are
measured in bins: horizontal bars denote the widths of the bins, and
the vertical bars denote 1$\sigma$ errors. The solid line in the
bottom panel shows the total flux curve (scaled by 10\%) for
comparison with the polarized flux. For comparison, 
the point spread function has a FWHM 
of $\approx$1\arcsec.  Note that the percentage
polarization increases significantly toward the NW, and that the
polarization position angle is roughly perpendicular to the major axis.  }
\label{impol} 
\end{figure}

The results of our one-dimensional imaging polarimetry are presented in
figure~\ref{impol}, and clearly demonstrate that the polarized flux is
spatially extended. In the central regions of the galaxy, the percent
polarization is $\sim$10\%, and rises to $\approx$23\%$\pm$4\% in the
low surface brightness `fuzz' in the NW portion. Note that since the
narrow emission lines from the galaxy that contaminate this large wavelength
range are probably unpolarized (see below), 
the true percentage
polarization of the continuum is slightly higher. In the central
regions, the total emission line equivalent width (in the observed
frame) is $W_{lines}^{obs}\approx$450\AA, and the true polarization is therefore larger 
by a factor of $(1-W_{lines}^{obs}/\Delta\lambda)^{-1}\approx 1.2$ and is 
likely to be $\sim$12\%. In the low surface brightness `fuzz', however,
the total emission line equivalent width is only $\approx$150\AA, and
the continuum polarization is insignificantly higher.  In all the
sampled bins the electric vector is roughly perpendicular to the major
axis of the galaxy. 

The bottom panel shows a comparison between the spatial variation of
the polarized flux (points) and that of the total flux (solid line).
The polarized flux has a somewhat flatter spatial distribution than the
total flux, primarily because of the excess polarization in the NW
`fuzz' region of the galaxy.  The excess polarization in this NW `fuzz'
may indicate less dilution of the continuum polarization by an
unpolarized component outside the main body of the galaxy.  The lower
polarization in the brighter regions of the galaxy relative to the NW `fuzz' 
may also be due to
geometric effects, resulting from the seeing smoothing over a
region where the electric vector position angles are changing
significantly thereby diluting the measured fractional polarization.
Alternatively, the scattering may not always take place along the
same direction relative to our line of sight; \eg, in the case of dust
scattering, it is conceiveable that the near side and far side may have
different fractional polarizations.  However, neither of these
possibilities explain the crude anticorrelation between surface
brightness and fractional polarization, which is naturally explained by
dilution by an extra unpolarized component.

We also computed the percentage polarization in the pure
continuum region $\lambda\lambda$5440--6515\AA\ (rest frame
1926--2307\AA). In two apertures of widths 2\farcs1 and 4\farcs2, both
centered on the centroid of the continuum emission, the percentage
polarization was measured to be 9.8\%$\pm$0.9\% and 10.9\%$\pm$0.9\%
respectively. If the polarized component were unresolved, the percent
polarization in the larger aperture should be $\approx$6.5\%$\pm$0.7\%,
since the total flux in the larger aperture is $\approx$1.5 times the
flux in the smaller aperture.  This measurement is therefore discrepant from the
prediction of the point source hypothesis by 4$\sigma$.

\subsection {Spectropolarimetry of the Continuum and Line Emission}

Rather than estimate the polarization in each resolution element, we
chose to first rebin the spectropolarimetric data into continuum and
emission line regions in order to optimize the signal-to-noise ratio of
the polarimetry.  The linear Stokes parameters and the errors were then
calculated for each spectral bin according to the formalism described
by Miller, Robinson, \& Goodrich (1988). 

\begin{figure} 
\plotfiddle{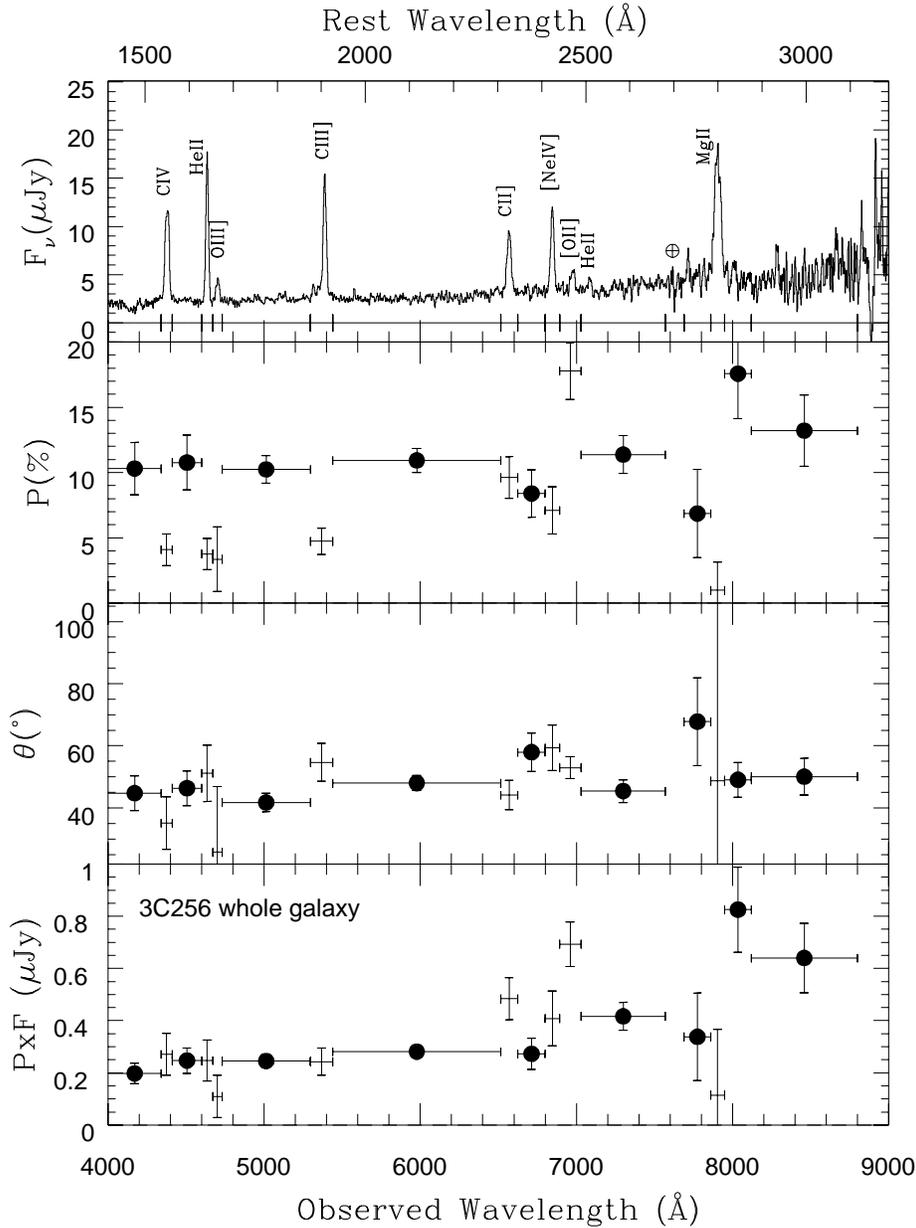}{7.0in}{0}{70}{70}{-230}{5}
\caption{The wavelength dependence of the total flux density $F_\nu$,
percentage polarization $P$, polarization position angle $\theta$, and
polarized flux $P\times F$ for 3C256 measured in an
4\farcs3$\times$1\arcsec\ aperture oriented along the major axis of the
galaxy. The abscissa is labelled by observed wavelength on the bottom
and rest frame wavelength on the top.  $P$, $\theta$ and $P\times
F_\nu$ are measured in bins: horizontal bars denote the widths of the
bins, and the vertical bars denote 1$\sigma$ errors. Bins with emission
lines are denoted by crosses and those dominated by continuum
emission are represented by solid circles. The region which has been
corrected for telluric A-band absorption feature is denoted by
$\oplus$. The continuum polarization is roughly constant across the
spectrum ($P\approx11\%$) and the narrow emission lines are
intrinsically unpolarized and dilute the underlying continuum
polarization.}
\label{3c256dat1pft} 
\end{figure}

\begin{figure} 
\plotfiddle{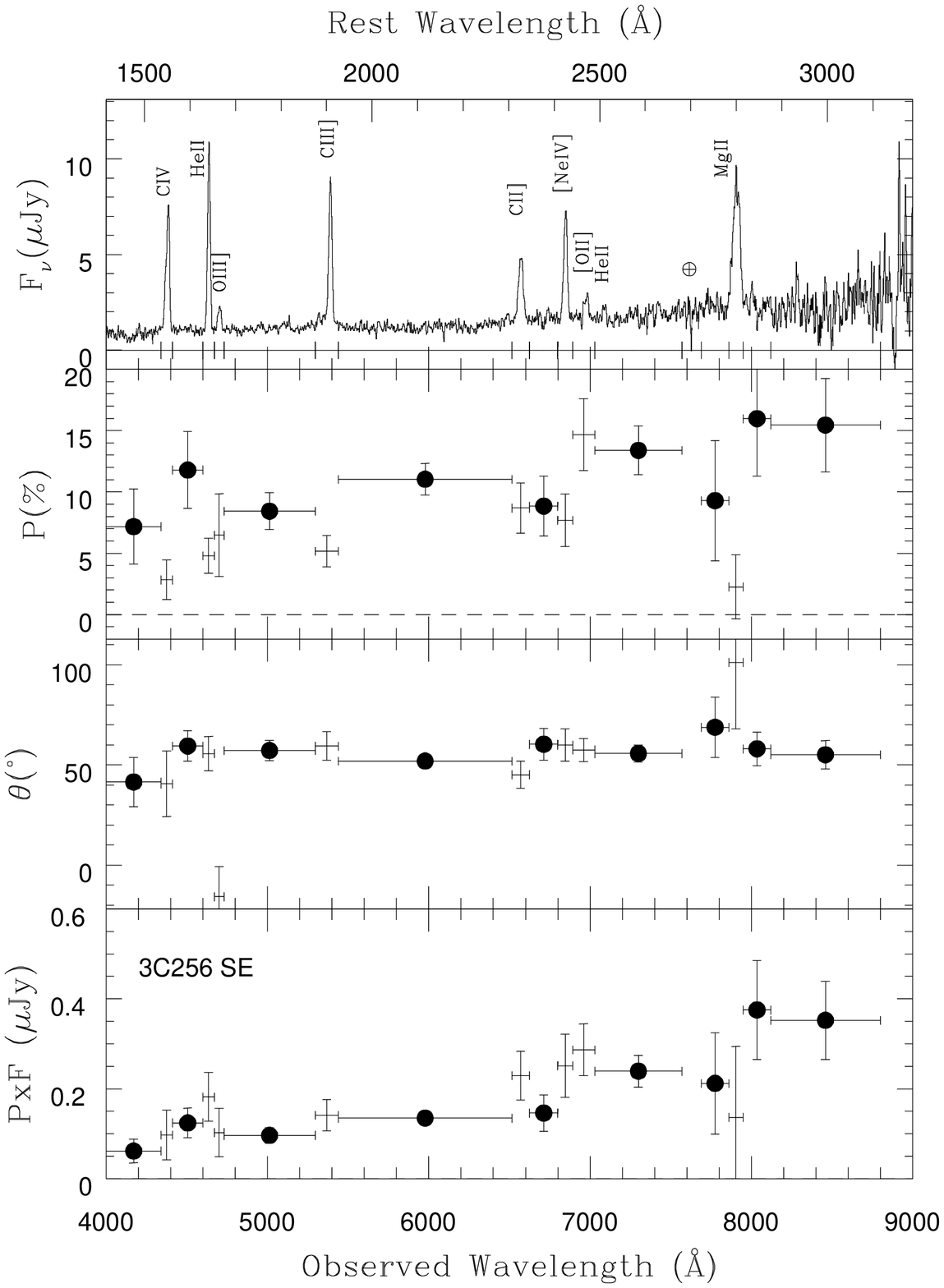}{7.0in}{0}{80}{80}{-230}{-80}
\vspace{0.5in}
\caption{Same as Fig.~5,  but for an aperture of size 
2\farcs1$\times$1\arcsec\ centered on the SE region of 3C256.}
\label{3c256dat4pft} 
\end{figure}

\begin{figure}
\plotfiddle{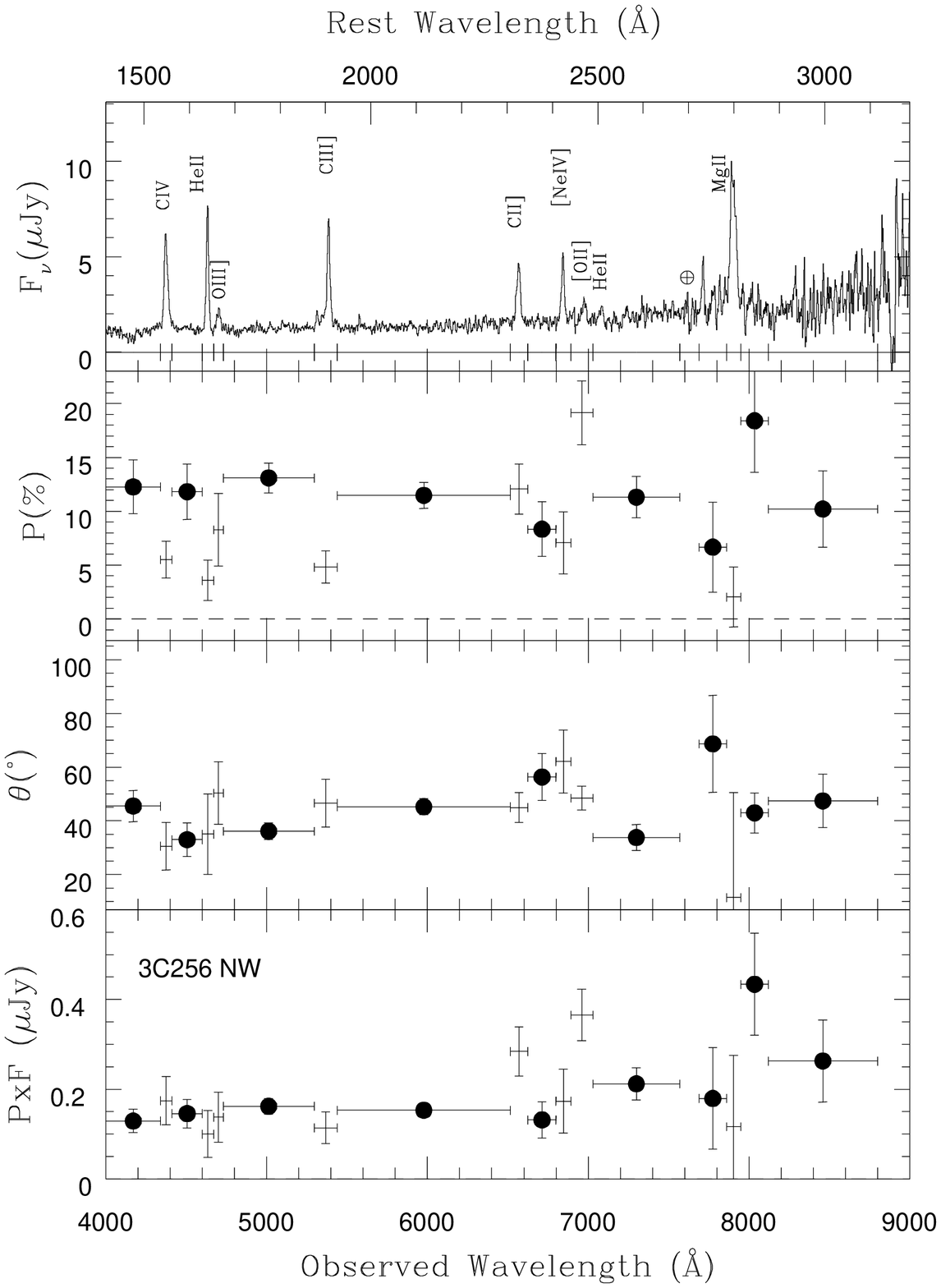}{7.0in}{0}{80}{80}{-230}{-80}
\vspace{0.5in}
\caption{Same as Fig.~5, but for an aperture of size 
2\farcs1$\times$1\arcsec\ centered on the NW region of 3C256.}
\label{3c256dat2pft}
\end{figure}

\begin{figure} 
\plotfiddle{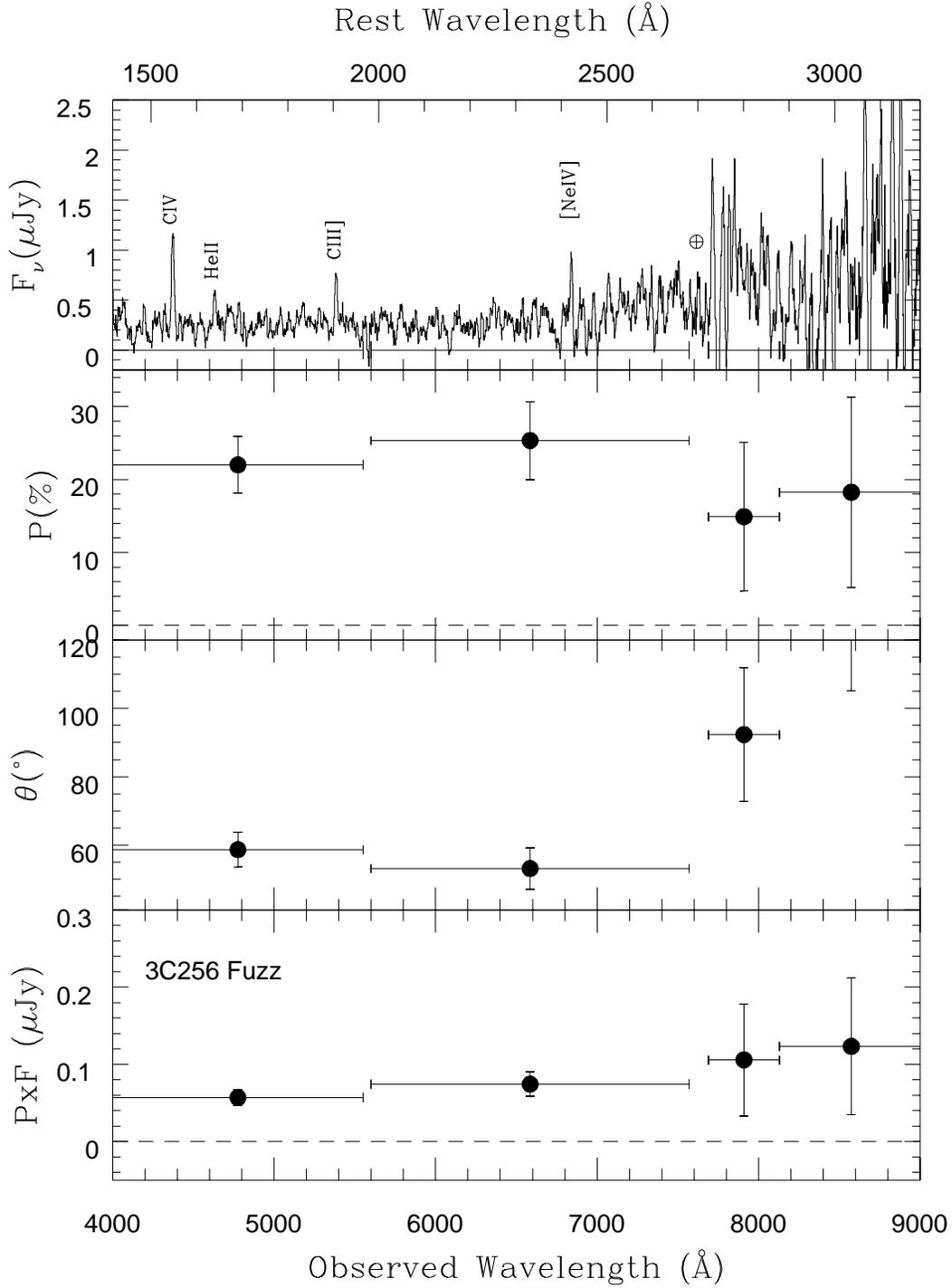}{7.0in}{0}{80}{80}{-230}{-80}
\vspace{0.5in}
\caption{Same as Fig.~5,  but for an aperture of size 
2\farcs1$\times$1\arcsec\ centered on the fuzz in the NW region of 3C256.}
\label{3c256dat5pft} 
\end{figure}

The results for the continuum polarization are listed in Table~2 for
three different apertures: one wide extraction (20 pixels or
4{\farcs}1) centered on the galaxy, and two smaller extractions (10
pixels or 2{\farcs}1) centered on the bright NW and SE continuum
emitting regions of the galaxy (see figure \ref{3c256bvr}; regions `a'
and `b' respectively of LeF\`evre \etal 1988).  In addition, we also
extracted the spectra of the faint `fuzz' in the NW region of the
galaxy (figure~\ref{3c256bvr}; region `c' of LeF\`evre \etal 1988). For
each spectral extraction, we tabulate the percentage polarization $P$,
an unbiased estimate of the percentage polarization $P_{unb}$
($=\sqrt{P^2 - \sigma_P^2}$; Wardle \& Kronberg 1974, Simmons \&
Stewart 1985), the error in the percent polarization $\sigma_P$, the
polarization position angle $\theta$ (of the electric vector), and the
error in the position angle $\sigma_\theta$. Bins that contain strong
emission lines or are contaminated by telluric absorption are flagged
in the last column. Electric vector position angles derived from $<2\sigma$ 
polarization 
measurements are highly uncertain, and are so noted by parentheses. 
We exclude the region around the A-band
(7570\AA\ to 7690\AA).  Figures \ref{3c256dat1pft} through
\ref{3c256dat5pft} show the variation of the percentage polarization
($P$), the polarization position angle ($\theta$), and the polarized
flux ($P\times F$) for the four different spectral extractions of
3C256. Since the `fuzz' in the NW region of the galaxy is very faint,
we were only able to estimate its polarization in very large spectral
bins (figure~\ref{3c256dat5pft}).

Table~3 lists the emission line polarizations for 3C256
measured in the wide (4{\farcs}1$\times$1\arcsec) aperture.  The
polarization of the narrow emission lines was estimated by first
subtracting a 3rd order polynomial fit to the continuum and then
calculating the linear Stokes parameters in bins centered on the
emission lines.  Within the errors, the narrow emission lines 
are all unpolarized.

The spatially extended continuum emission from 3C256 is polarized at
the $\approx$11\% level.  The electric vector of the polarized light is
oriented at PA$\approx$49\deg$\pm$5.6\deg, which is roughly orthogonal
to both the major axis of the optical emission and the radio axis of
the galaxy.  Within the observational errors, the percentage
polarization and the polarization position angle are both roughly
constant with wavelength in all parts of the galaxy. It should be
noted, however, that the percentage polarization in the SE component 
(figure~\ref{3c256dat4pft})
shows marginal evidence for both an increase with wavelength
(1.7$\sigma$) and a different polarization position angle compared to the NW
component (1.5$\sigma$) which may imply that the scattering
population, the diluting radiation and the scattering geometry 
are different in this region of the
galaxy, but higher signal-to-noise ratio data are required to confirm
this.  The narrow emission lines are unpolarized and dilute the
underlying continuum polarization.  In all the extractions, there is
very marginal evidence for increased polarization in the continuum band
situated in the red wing of the MgII emission line.


\section {Discussion}
\subsection{Scattered Nuclear Light from 3C256} 
\nopagebreak
The relatively high percentage polarization of the continuum emission and
the orientation of the polarization position angle imply that a
significant fraction of the observed extended continuum emission is
scattered light from a hidden nuclear source. The exact nature of the
nuclear source is unknown at this juncture, but we speculate that it is
most likely to be the AGN powering the radio source.
Since the percentage polarization is roughly constant with wavelength in 
all parts of the galaxy (with the possible exception of the SE 
region; see figure~\ref{3c256dat4pft}), the
spectrum of polarized flux therefore has the same spectrum as the total
flux spectrum from the galaxy. Under the assumption that the total flux
spectrum is a power law $F_\nu\sim\nu^{\alpha}$, then both $F_\nu$ and
$P\times F_\nu$ have optical spectral indices of $\alpha\approx -1.1\pm0.1$, 
which is fairly typical of quasar spectra in this wavelength region.

Although the continuum emission may be dominated by scattered AGN
light, the narrow emission lines are not polarized and therefore
probably dominated by {\it in situ} emission.  The total light 
spectrum shows that the MgII$\lambda$2800 emission line is slightly 
broader in the SE part of the galaxy, but this is probably due to the 
kinematics of the narrow line emitting gas. Higher
signal-to-noise ratio data are required to accurately determine the
spectrum of the scattered light in this region.

The hypothesis that the radio galaxy 3C256 contains a buried quasar is
convenient from the perspective of AGN unification ideas, which predict
that FRII radio galaxies and steep spectrum radio loud quasars are
intrinsically similar objects, their observed differences being
primarily due to the effects of orientation and dust obscuration (see
Antonucci 1993 for a review).  If the nuclear source in 3C256 is indeed
a quasar, the scattered light spectrum should show the broad emission
lines of CIV, CIII] and MgII that typify quasar spectra (\eg, Antonucci
\& Miller 1985, Miller \& Goodrich 1990).  We attempted to fit the
MgII$\lambda$2800 and CIII]$\lambda$1909 features in the total
light spectrum with a narrow gaussian component and a broad gaussian
component centered at the same wavelength. Within the range of FWHM
normally observed in quasars (${\rm 2100\ km\ s^{-1} \simlt FWHM_{\rm
MgII, QSO} \simlt 10300\ km\ s^{-1}}$; Brotherton \etal 1994), we
derive an upper limit to the equivalent widths of the broad components of
$W_{rest}^{\rm MgII,CIII]} < 25$\AA. Observed equivalent widths of
the broad MgII feature in quasars range from 15\AA\ to 60\AA\ (\eg,
Steidel and Sargent 1991), and therefore this limit does not rule out
the `hidden quasar' hypothesis.  A similar limit is more difficult to
obtain for the CIV line since it is lies close to an absorption feature
in the spectrum (\S 4.2).  Unfortunately, the signal to noise ratio of
the present data is inadequate to determine any useful limits to the
presence of broad lines in the polarized flux spectrum.

\subsubsection{What is the Scattered Fraction?}

Before we discuss the scattering mechanism responsible for the polarization 
in 3C256, it is important to
determine the fraction of scattered light in the observed spectrum.
Since the percentage polarization is $\approx$11\% across the main 
body of the galaxy ({\it e.g.,} figure~\ref{impol}), we could naively
assume that the scattered light is completely polarized, and that
$\approx$89\% of the observed light is {\it in situ} diluting
(unpolarized) emission from the galaxy. However, this scenario is
extremely unlikely given the spectrum of the {\it unpolarized}
continuum emission: since the percentage polarization is wavelength
independent, the unpolarized flux would have the same spectrum as the
total light.  The remarkable similarity between the rest frame continuum spectra
of 3C256 and NGC1068 (figure~\ref{3c256vs1068}) suggests that the rise
in the continuum emission longward of $\lambda\sim 2300$\AA\ 
relative to a smooth extrapolation from shorter wavelengths may be
primarily due to the presence of blended FeII emission lines 
({\it e.g.,} Antonucci, Hurt, \& Miller 1994).  Since broad FeII emission
over such a large volume would be completely unprecedented, it is far
more likely that a significant fraction of the continuum emission is
scattered nuclear light.

A crude estimate of the true percentage polarization of the scattered
light (and the relative importance of the diluting continuum) is
provided by our one-dimensional imaging polarimetry. If we assume that
the true percentage polarization of the scattered light is roughly
constant across the face of the galaxy, then the rise in the percent
polarization in the region of the NW `fuzz' can be interpreted as a
decrease in the diluting flux in this region. As discussed above, in
the wavelength range $\lambda\lambda_{rest}$1416--2300\AA\ the percent
polarization in the extended NW `fuzz' region is 23.0\%$\pm$3.7\%
whereas it is only 9.6\%$\pm$0.9\% in the central region. If we assume
that the intrinsic polarization $P_{\circ}$ of the scattered light is
constant over the entire galaxy, then $P_\circ\ge23\%$ since the
intrinsic percentage polarization must be at least equal to that observed
in the NW `fuzz'.  Since the observed polarization is given by
$P=P_\circ F_{scat}/(F_{scat}+F_{dil})$ where $F_{scat}$ is the
scattered flux and $F_{dil}$ is the unpolarized diluting continuum
emission, the fraction of diluting flux in the central regions is $90\%
\simgt ({{F_{dil}}\over{F_{tot}}})\simgt 58\%$.  Moreover, since the
total flux in the `fuzz' is $\approx$1/6 that in the center, the
scattered flux in the `fuzz' must be only $\approx$0.4 that in the
central region. Similar continuum components that dilute the continuum 
polarization have also been detected in nearby Seyfert galaxies (\eg, 
Antonucci \& Miller 1985, Miller \& Goodrich 1990, Tran 1994$a,b,c$).

The estimate that the diluting radiation contributes at least 58\% of
the total light in the central regions is perhaps somewhat surprising.
As mentioned above, this implies that the continuum spectrum of the
unpolarised radiation must be similar to that of the total spectrum,
and then its resemblance to the nuclear spectrum of NGC1068 is
puzzling. Of course, our simple minded assumption that the intrinsic
polarization is constant throughout the galaxy may be false: $P_\circ$
might be larger in the `fuzz' if the scattering mechanism is different,
or enhanced due to an increased population of scatterers.  It is also
important to note that the polarization in the brighter central regions
of the galaxy may be diluted by various geometric and sampling effects,
as discussed in \S 3.2.  High spatial resolution imaging polarimetry is
required in order to properly address this question.

\subsubsection{What is the Scattering Mechanism?}

Is it possible to determine from these data whether the scatterers are
primarily electrons or dust?  Electron scattering is wavelength
independent over the entire spectral regime. Dust generally produces
wavelength dependent scattering at optical rest frame wavelengths, but
in the UV the spectrum produced by dust scattering depends upon the
albedo of the dust grains, which in turn depends upon the physical
properties of the grains (composition, size distribution).  For
example, the dust cloud in PKS~2152$-$69 is optically thin at optical
wavelengths and causes Rayleigh scattering of the optical AGN light,
but the scattered flux spectrum flattens at UV wavelengths (di Serego
Alighieri \etal 1988, Fosbury \etal 1990).  In fact, at UV wavelengths
optically thick dust scattering can mimic the wavelength independence
of electron scattering for wavelengths longward of about
2500\AA\ (Kartje 1995; Laor, pers.~comm.).

In this section, we use mass estimates for the scatterers to try to
distinguish between electrons and dust as the dominant scattering
population. We first discuss a simple model for electron scattering,
then consider whether hot ($T>10^7$~K) or warm ($T<10^6$~K) gas is
responsible for scattering the light, and finally briefly discuss the
possibility of dust particles as the scattering population.

\begin{figure}
\plotfiddle{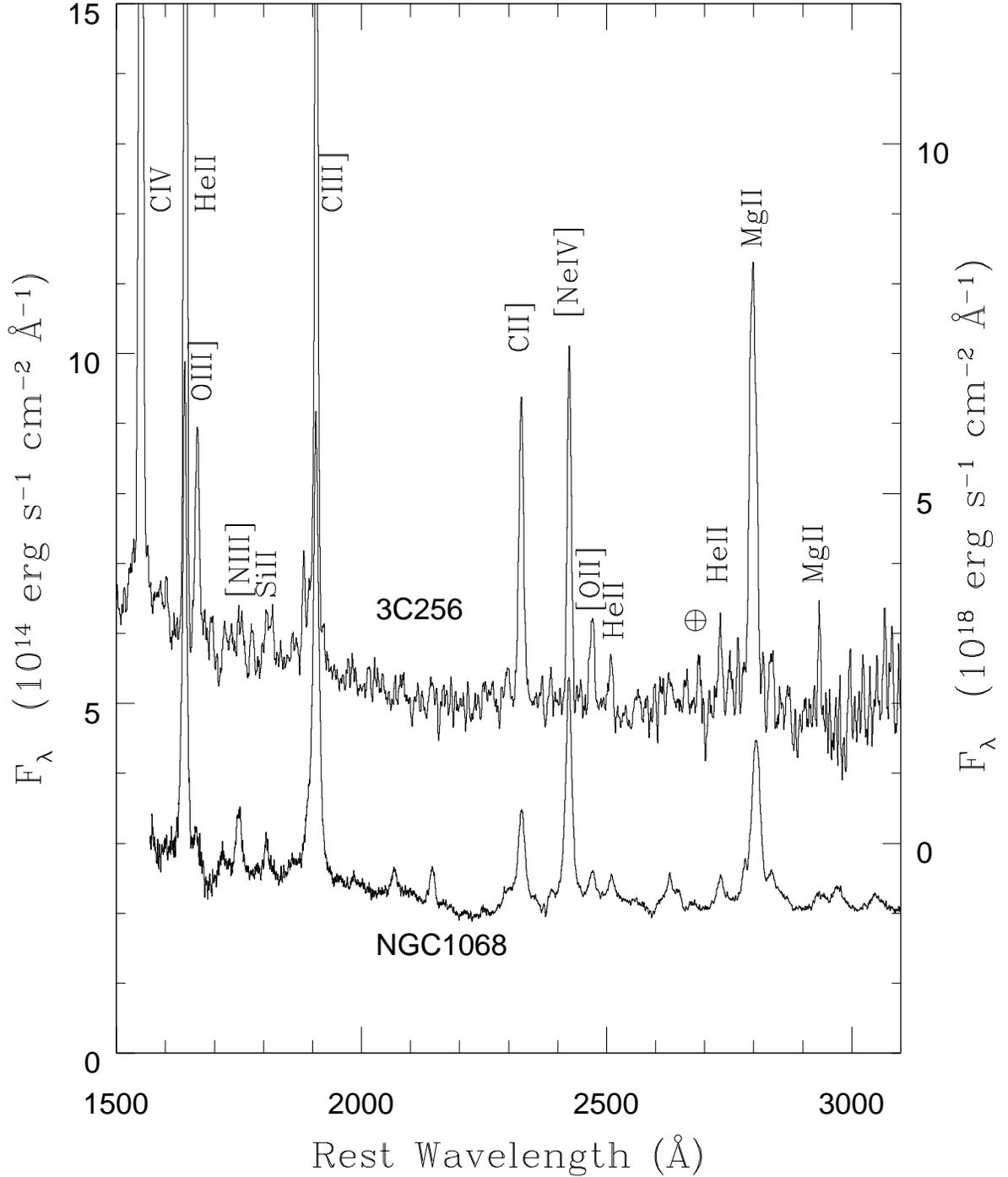}{7in}{0}{100}{100}{-260}{-80}
\caption{Comparison of the rest frame UV continuum spectra of 
3C256 and NGC1068 (from Antonucci, Hurt, \& Miller 1994). 
The ordinate is labelled with the flux density scales for NGC1068
and 3C256 on the left and right axes respectively.
Note that the spectrum of 3C256 is sampled in 
a 43~kpc $\times$ 11.3~kpc aperture (3{\farcs}8$\times$1\arcsec) 
whereas the spectrum of NGC1068 is from a nuclear 
region of size 471 pc $\times$ 153 pc (4{\farcs}3$\times$1{\farcs}4).}
\label{3c256vs1068}
\end{figure}

One intriguing piece of evidence is provided by the total light
spectrum of 3C256. Figure~\ref{3c256vs1068} shows a comparison of the
rest frame UV spectra of 3C256 and NGC~1068 (from Antonucci, Hurt and
Miller 1994). It is important to note that the spectrum of NGC1068 was
obtained in a 500~pc $\times$ 160~pc 
aperture centered on the nucleus, whereas the spectrum 
of 3C256 is extracted in an aperture $\approx 45\times 11$~kpc in size.  
In NGC1068, the percentage polarization in the nucleus is wavelength
independent for at least ${\rm 1500\AA < \lambda < 9000\AA}$ (after
correction for dilution by starlight; Miller \& Antonucci 1983,
Antonucci \etal 1994) and is therefore believed to be indicative of 
electron scattering. Although the similarity between the spectra
displayed in Fig.~\ref{3c256vs1068} may suggest, by analogy, that
electrons are also responsible for scattering and polarizing the light
in 3C256, the inefficiency of electron scattering argues against 
this mechanism as the origin of the scattered light due to the implied
large mass of ionized gas. 

\noindent\underbar{\it A Simple Biconical Electron Scattering Model for 3C256}

Let us consider the null hypothesis that {\it all} the observed
optical/UV flux from 3C256 is due to electron scattering of nuclear
light. Let us further assume that the nuclear source in 3C256 radiates
into two cones of half opening angle $\theta_{cone}$, but is hidden
from direct view (\ie, $\theta_{cone}<\theta_{los}$, where
$\theta_{los}$ is the angle between the line of sight and the cone
axis), and that the cones are filled with ionized gas of uniform
electron density $n_e$.  For the axisymmetric geometry, uniform density
and optically thin case considered here, the two angles $\theta_{cone}$
and $\theta_{los}$ are related via the intrinsic percentage
polarization (Brown \& McLean 1977; also see equation 3 of Miller, Goodrich \&
Matthews 1991).  In particular, if the intrinsic polarization $P_\circ$
is greater than 23\% ({\it i.e.,} at least as large as the observed
polarization of the NW `fuzz'), then $\theta_{cone}$ is always less
than $\theta_{los}$.  This model is similar to that presented for
NGC~1068 by Miller, Goodrich and Mathews (1991), and we refer the
reader to that paper for details regarding these arguments.

In the case of 3C256, the opening angle for the scattering cone is
probably not large: the optical continuum and line emission of the
galaxy are confined to a fairly narrow, elongated structure
(figure~\ref{3c256bvr} and LeF\`evre \etal 1988). If the gas
distribution is assumed to be spherically symmetric but only
illuminated/ionized within the cone, the optical image can be used to
constrain the opening angle to the range
$15\deg\simlt\theta_{cone}\simlt30\deg$. This is slightly less than the
value of $\approx44\deg$ argued by Barthel (1989) to be the typical 
cone opening angle, and is within the range of cone angles
observed for ionization cones in low redshift Seyfert galaxies (Wilson
\& Tsvetanov 1994).  We therefore adopt this range
($15\deg\simlt\theta_{cone}\simlt30\deg$) as representative for the
discussion that follows.

The ratio of the scattered luminosity to the incident luminosity is 
$$
\eta\equiv {{L_{scat}}\over{L_{inc}}} = \int\int \phi(\theta_{scat})\sigma n_e dr\ d\Omega
$$
where $\phi(\theta_{scat}) = {{3}\over{16\pi}}(1 + {\rm
cos^2}\theta_{scat})$ is the scattering phase function for electron
scattering, $\sigma_T$ is the Thomson cross section, and the integrals
are performed over the path length and solid angle occupied by the
scattering region.  For the conical model adopted here
$$
\eta \approx 0.015 n_e R_{20} f(\theta_{los},\theta_{cone})
$$
where $R_{20}$ is the length of each cone in units of 20~kpc, and 
$$
f(\theta_{los},\theta_{cone}) = (3-\mu_{los}^2)(1-\mu_{cone})+{{1}\over{3}}(3\mu_{los}^2-1)(1-\mu_{cone}^3)
$$
where $\mu_{cone}={\rm cos}\theta_{cone}$ and $\mu_{los}={\rm
cos}\theta_{los}$ (Miller, Goodrich, \& Matthews 1991, Brown \& McLean
1977). 
Since the optical depth $\tau\propto n_eR_{20}$, as long as $\eta<0.027$ and $\theta_{cone}>15\deg$, the optical depth is $\tau < 2/3$ and 
the scattering can be considered to be optically thin. 

Hence, if we know $\eta$ we can derive the electron density required to
produce the observed scattered flux.  The total mass in the ionized
component of the ISM (responsible for the electron scattering) {\it
within the cone} is then 
$$ 
M_{ism}^{cone} \approx 5.4\times 10^{13} \eta ({{1 - {\rm cos}\theta_{cone}}\over{f(\theta_{los},\theta_{cone})}})R_{20}^2 \msun .  
$$ 
The derived mass is fairly insensitive to the exact value of the cone
opening angle.

What is the value of $\eta$?  The radio loud quasars in the redshift
range $1<z<2$ from the Steidel and Sargent (1991) sample have rest frame
2200\AA\ luminosities in the range $46.4 < {\rm log[\nu
L_\nu(erg\ s^{-1})]} < 47.6$.  The total luminosity at 2200\AA\ of
3C256 is
$${\rm log}L_{scat} = {\rm log}[4\pi d_L^2\nu f_\nu] \approx 45.14 + {\rm log}\left({{f_\nu}\over{7\mu{\rm Jy}}}\right)$$
so if
$$ \eta = 0.014\left({{L_{inc}}\over{10^{47}{\rm erg\ s^{-1}}}}\right)^{-1}$$
then, for $15\deg < \theta_{cone} < 30\deg$, 
$$ 8.3\left({{\eta}\over{0.014}}\right){\rm cm^{-3}} \simgt n_e \simgt 2.2\left({{\eta}\over{0.014}}\right){\rm cm^{-3}}$$
and 
$$ M_{ism}^{cone} \approx 2.4\times 10^{11}R_{20}^2\left({{\eta}\over{0.014}}\right)\msun$$
Note that this is only the mass of the ionized gas responsible for the 
scattering ({\it i.e.}, within the cone); 
if we assume that the gas is in a spherical distribution, the 
mass estimates for the ionized component would be larger by $4\pi/\Omega = 
(1-\mu_{cone})^{-1}$, or 
roughly one order of magnitude. 

The density and mass estimates above are derived based on the
assumption that {\it all} the light observed from 3C256 is due to
electron scattering. This assumption is at odds with the discussion in
the previous section, where, based on the assumption that the intrinsic
polarization is constant across the galaxy, the scattered light in the
central regions of the galaxy was suggested to comprise less than 42\% of
the total light. Including a significant contribution from an
unpolarized, diluting continuum alters the above arguments in two
ways.  Since the intrinsic polarization is higher than the measured
polarization, the scattering angle is larger, and the cone angle is
constrained to a smaller range.  This reduces the lower limit on the
electron density, but does not affect the mass estimate.  The more
important effect is that the estimate of $L_{scat}$, and therefore
$M_{ism}^{cone}$ decreases by a factor of ($11\% / P_{\circ}$).

Under the null hypothesis, the lower limit on the mass of the ionized
gas in the cone is comparable to the total mass of our Galaxy. This
does not necessarily rule out electron scattering, however, but implies
that the ionized component {\it alone} has a mass comparable to that of 
the baryonic mass of a giant elliptical galaxy. 
If we restrict the total mass of the 3C256 to be less than that of a
present day cD galaxy ($\approx 10^{12}$\msun), this raises the
question as to why most of the mass is confined to the region within
the ionization cone (or the boundaries defined by the radio source).

\noindent\underbar{\it $T>10^7$~K Electrons as the Scattering Population}

A possible explanation for such a large mass in ionized gas is to invoke
the suggestion of Fabian (1989) that radio galaxies may be surrounded
by large halo of hot ($T\approx 2\times10^7$~K) gas, similar to that
observed to surround most central cluster galaxies. The average 
gas density within 20~kpc of 2.2 -- 8.3 \cm3\ required by the scattering 
hypothesis is larger than that
observed in most clusters, but is similar to the density in the model adopted by
Fabian (1989). If the hot electrons are confined (for some unknown reason) 
to a spherical region within 20~kpc of the nucleus, the X-ray flux in the 
observed energy range 0.1--2.4~keV is 
$$
f_X = {{1}\over{4\pi d_L^2}}\int_{0.1(1+z){\rm keV}}^{2.4(1+z){\rm keV}}\int_V \varepsilon_\nu d\nu dV 
$$
where $\varepsilon_\nu$ is the emissivity of thermal bremsstrahlung. 
Assuming that the density is constant over this region 
implies that 
$$
f_X \approx 2.7\times 10^{-12} R_{20}^3 \left({{n_e}\over{5\cm3}}\right)^2 T_7^{1/2} \left( e^{-0.33/T_7} - e^{-7.87/T_7} \right) {\rm erg/s/cm^2}
$$
or, in the R\"osat PSPC detector (assuming $N_H = 1.1\times 10^{20}$; 
Crawford \& Fabian 1995), this corresponds to 
$$
\approx 0.3 R_{20}^3 \left({{n_e}\over{5\cm3}}\right)^2 T_7^{1/2} \left( e^{-0.33/T_7} - e^{-7.87/T_7} \right) {\rm cts/s}.
$$
Hence the emission from even the very central 20~kpc is nearly 200
times greater than the upper limit to the soft X-ray flux from 3C256 of
$f_X<0.00174\ {\rm cts/s}$ quoted by Crawford \& Fabian (1995). If we
further assume that the hot gas has a distribution similar to that
observed in most clusters ($\rho\propto (1+r^2/r_c^2)^{-1}$,
$r_c\approx0.2$Mpc), the predicted flux and count rate exceed the upper
limit by a factor of $>10^4$! In addition to this problem, the cooling
time of the gas would be very short ($t_{cool}< 5.4\times 10^6$yr), and
the mass deposition rate would be extremely high.  

The only possibility
of retaining hot electrons as a feasible scattering population in the
case of this high redshift radio galaxy is to contrive that the
electrons are completely restricted to the region within the cone, with
negligible emission measure outside the radio source boundary, {\it and} to
have the central source be among the intrinsically most luminous
quasars in the universe. If we reverse the above argument and use the
R\"osat PSPC upper limit to predict the number density of the hot
electrons, we find that scattering by this hot population does not
contribute significantly to the observed polarized flux.  Broad
MgII$\lambda$2800\AA\ line emission with widths similar to that seen in
quasars is observed in the polarized flux spectra of the $z\approx 0.8$ 
radio galaxies 3C226 and 3C277.2 (di Serego Alighieri \etal 1994) and 
the $z=1.206$ radio galaxy 3C324 (Cimatti \etal 1996), and in the 
total light spectrum of the $z=0.811$ radio galaxy 3C265 (Dey \& Spinrad 1996) 
and rule out hot electrons as the scattering population for these galaxies
as well.

\noindent\underbar{\it $T<10^6$~K Electrons as the Scattering Population}

Although we can convincingly rule out hot electrons as the dominant
scatterers, it is more difficult to rule out scattering by a large
population of cooler ($T<10^6$~K) electrons.  In fact, if we assume that
the cones are filled with $T=10^4$~K,  $n_e=5$~\cm3\ gas, then the Ly$\alpha$
luminosity produced due to recombination would be $\approx 5.4\times
10^{44}\ {\rm erg\ s^{-1}}$ (Osterbrock 1989). 
This is not very different from the observed
Ly$\alpha$ luminosity of $\approx 2\times 10^{44}\ {\rm erg\ s^{-1}}$
measured by Spinrad \etal (1985), and implies that the same gas
responsible for the line emission could also be responsible for
scattering the nuclear emission. 

The main limitation to the present discussion of whether electrons can
be responsible for scattering the light is the lack of a reliable,
independent density estimate. At present, we can only very crudely constrain the
electron density in a one--zone model 
to lie in the range $10^5\ {\rm cm^{-3}} \simgt n_e \simgt
2\ {\rm cm^{-3}}$, where the upper limit is the critical density for
the collisional deexcitation of the [NeIV]$\lambda$2424 emission line
(Spinrad \etal 1985), and the lower limit is based on the electron
scattering arguments outlined above. Modelling of the emission line
spectrum should provide better constraints to the density, temperature
and filling factor of the gas, and we will investigate this possibility
in more detail in a future paper.

If the observed light is scattered primarily by electrons, the dust--to--gas 
ratio in the scattering cones must be very small. Dust particles
are much more efficient than electrons at scattering light: for grains
of radius 0.1$\mu$m and mass density 3${\rm \ g\ cm^{-3}}$ embedded in
a pure hydrogen nebula, the relative efficiency of dust scattering to
electron scattering is $$ {{\epsilon_{dust}}\over{\epsilon_{gas}}} \sim
377\left({{M_{dust}/M_{gas}}\over{0.006}}\right)$$ where 0.006 is the 
Galactic dust--to--gas ratio, and we have
assumed that the scattering cross section of a dust grain is identical
to its geometrical cross section, and the ratio of the scattering phase
functions of dust and electrons is of order unity. Therefore, if the
scattering is dominated by electrons, the dust--to--gas ratio in the
scattering cones must be less than $10^{-3}$ that in our galaxy,
implying a dust mass (in the cones) of less than $4\times 10^6
(n_e/5{\rm cm^{-3}}) R_{20}^3$\msun. The low dust--to--gas ratio and
dust mass may imply that the radiation within the cone is 
responsible for destroying the dust grains, or that the thermal history of 
the gas has prevented significant grain formation.

\noindent\underbar{\it Dust as the Scattering Population}

Although dust particles are much more efficient than electrons at
scattering light, their polarization efficiency can be lower than that
of electrons. In addition multiple scattering and variations in the
grain properties can result in a high scattered fraction with
relatively low percentage polarization in the UV as is often observed
in Galactic reflection nebulae.  
Moreover, model calculations by Laor (1995) suggest that the increased
scattered flux observed in 3C256 for $\lambda_{rest} \simgt 2300$\AA
\ ($\lambda_{obs} \sim 6500$\AA) which we interpreted as resulting from
FeII emission in the preceding discussion, can also be produced by dust
scattering (Laor, pers.\ comm.).  If we attribute all the observed
continuum emission to dust scattering off grains with typical radius of
0.1$\mu$m and mass density 3${\rm \ g\ cm^{-3}}$, the required mass of
dust within the cones is $$M_{dust}\sim 10^8 R_{20}^2
\left({{L_{inc}}\over{10^{47}{\rm erg\ s^{-1}}}}\right)^{-1}\Msun$$
where we have assumed that the scattering angle is $\approx90\deg$, and
the Henyey--Greenstein (1941) scattering asymmetry parameter is $g=1/2$
(\eg, Savage \& Mathis 1979 and references therein). This estimate for
the dust mass, like the mass estimate for the ionized gas resulting
from the electron scattering assumption, is nearly 10 times larger than
the mass of dust in the disk of our Galaxy.

This crude mass estimate by itself does not rule out dust scattering as
the mechanism, since we know little about the large scale dust and gas
properties of these high redshift galaxies. However, the large amount
of dust predicted by this simple picture, if distributed fairly uniformly, 
would result in significant
extinction which should be apparent in the flux ratios of the narrow
emission lines.  If the dust is uniformly distributed within the cone,
the optical depth through the cone perpendicular to the cone axis at a
distance of 10~kpc from the center is $\tau_{dust} \approx 1$, which
corresponds to $E(B-V) \approx 0.4$.  However, the reddening must be
low, since the HeII$\lambda$2733/HeII$\lambda$1640 ratio is observed to
be $\approx$0.039, which is close to the Case B recombination value of
0.032 ($T=10^4$~K, Hummer \& Storey 1987). 

This argument against dust scattering is only suggestive, since the
dust may be distributed differently from the line emitting gas. It is
entirely plausible that the dusty, scattering medium is clumped with
high density and low filling factor, and that the line emission arises
from a more diffuse, high filling factor inter--clump medium and is
therefore not strongly affected by dust extinction or reddening. Moreover, 
at large dust optical depths, the scattering efficiency can compensate 
for the reddening, resulting in low measured extinction (Laor, pers.
comm.). In addition, some preliminary modelling of the broad band rest
frame UV to optical spectral energy distribution of 3C256 kindly
provided to us by Dr.\ S.\ di Serego Alighieri suggests that a more
sophisticated three component model of a dust scattered quasar, single
age old stellar population, and nebular continuum is also able to fit
the observed spectrum fairly well.  However, the critical tests of
whether dust scattering is the primary mechanism responsible for the
observed properties of the galaxy would be near-infrared polarimetry
(the dust scattering model predicts much lower fractional polarization
in the near-IR) and near-infrared spectroscopy (to search for the
spectral signatures of an old population).

Another (albeit more speculative) argument against dust as the dominant 
scattering mechanism in 3C256 follows from the discussion in \S4.3 where 
we argue that 3C256 has not yet formed the bulk of its stars. If dust forms
only in cool stellar envelopes, it is unlikely that dust production on a large 
scale in this young galaxy is efficient enough to produce the large 
mass of dust required for scattering. Dust formation in the ejecta of 
supernovae, however, may be more efficient in dispersing dust over a large 
volume. 

The primary argument for electrons and against optically thin dust as
the dominant scattering population is the lack of any wavelength
dependence to the observed fractional polarization. This is exactly
what one would expect for electron scattering, whereas dust scattering
should imprint some wavelength dependent signature on the scattered
light. For example, dust in Galactic reflection nebulae generally
produces a scattered light (and probably polarized flux) spectrum that
is bluer than the incident spectrum by $\nu^{0.8-1}$ (\eg, Calzetti
\etal 1995).  A particularly extreme example is that of the scattering
cloud in PKS~2152$-$69, which bluens the incident spectrum by $\nu^{4}$
(di Serego Alighieri \etal 1988, Fosbury \etal 1990). In addition,
recent scattering models of the UV continuum light in AGN (Kartje 1995)
have shown that although optically thick dust scattering can produce
wavelength independent fractional polarization for wavelengths $\lambda
\simgt 2500$\AA, it also usually produces a sharp rise in the
fractional polarization toward shorter wavelengths ($1000 - 2000$\AA).
This is not observed in 3C256, where $P$ is observed to be wavelength
independent down to at least 1500\AA.  Higher signal-to-noise ratio
spectropolarimetry of 3C256 in the optical (rest frame UV) and
near-infrared (rest frame optical) directed at identifying features in
the polarized flux (such as the 2200\AA\ bump) will eventually provide
a definitive test between dust and electron scattering.

\subsection {An Absorption Feature in the Total Light Spectrum}

\begin{figure}[b]
\plotfiddle{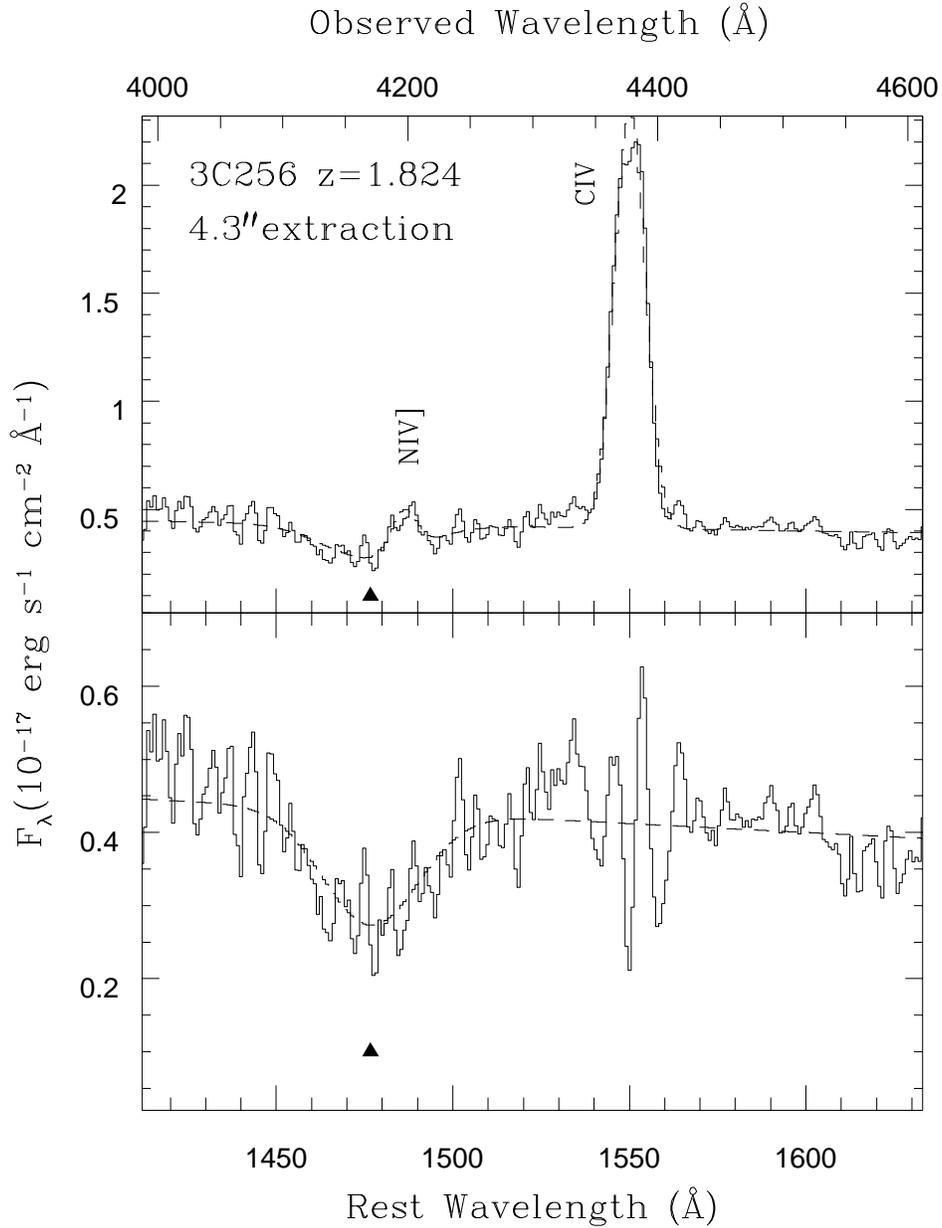}{7.0in}{0}{80}{80}{-230}{-80}
\caption{Detail of the coadded spectrum of 3C256 (extraction width
4\farcs3) showing the putative absorption feature at
$\lambda_{obs}\approx 4170$\AA ($\lambda_{rest}\approx 1477$\AA\ if the
absorption is at the same redshift as the narrow emission lines). The
abscissa is labelled with both the observed and 3C256 rest wavelength
scales on the top and bottom respectively.  The upper panel shows the
total spectrum overlaid with a fit to the continuum, the absorption
line and the NIV] and CIV emission lines.  The bottom panel shows the
same spectrum after subtraction of the NIV] and CIV emission lines. The
continuum is fit with a straight line and both the absorption and
emission lines are modelled as gaussians.}
\vspace{6truein}
\label{absline} 
\end{figure}

An intriguing characteristic of the total light spectrum is the
presence of a strong, broad absorption feature centered at
$\lambda_{obs}\approx 4170$\AA\ ($\lambda_{rest}\approx 1477$\AA;
figure~\ref{absline}). If the line has the same redshift as the narrow
line emission from 3C256, the rest wavelength is $\lambda_{rest}\approx
1477$\AA. This absorption feature is present in the coadded spectra of
both the upper and lower rays of the spectropolarimeter, and is located
in a spectral region which is not unduly affected by telluric lines or
features in the flat field. The feature is seen in the raw extracted
spectra of both the upper and lower rays, and can be seen in the
two-dimensional spectrum. In addition, no absorption feature at the
same observed wavelength is seen in the spectra of other objects
obtained on the same night, and it is therefore unlikely that the
absorption line is an instrumental artifact. In addition, the feature
is not produced in the flux calibration process: the flux calibration
sensitivity curves are smooth across this feature. An absorption line
at roughly the same wavelength ($\lambda_{rest} \approx 1480$\AA) was
also very marginally detected by Chambers and McCarthy (1990) in their
composite radio galaxy spectrum which was constructed by adding spectra
of 3C256 and 3C239.

The weak emission line near the center of the broad absorption trough
(figure~\ref{absline}) is most probably NIV]$\lambda$1486.  After
correcting for the NIV] line emission, the broad absorption has an
equivalent width in the observers' frame of $\approx$35$\pm$5\AA, and a
FWHM$\approx$6450$\pm$100\kms. If the line has the same redshift as the
narrow line emission from 3C256, the rest equivalent width is $\approx
12.4$\AA. Since at least 10--40\% of the continuum light is scattered
AGN continuum emission, the intrinsic equivalent width of this feature
is even higher ($\sim 14-21$\AA) if this feature arises in the {\it in 
situ} component alone.

What is the origin of this absorption? The possible interpretations of the 
absorption feature fall into three general categories according to whether 
it is present in the diluting component, the scattered component, or is 
imposed on the total spectrum by an external absorbing medium. These three 
situations may be distinguished by the behaviour of the percentage polarization
$P$ across the absorption feature: in the first case, $P$ should 
rise in the trough; in the second, $P$ should decrease in the trough; and 
in the third, it should remain constant across the feature. Unfortunately, 
the data presented in this paper are inadequate to conclusively distinguish
between these possibilities, but we consider the various options in more
detail below in anticipation of better data. 

\noindent\underbar{\it Stellar Absorption?}

If the absorption is a property of the unpolarized, diluting component,
then the most likely (and certainly most intriguing) possibility is
that it is a stellar photospheric line.  Absorption features in the
spectra of high redshift radio galaxies were first tentatively identified by
Chambers \& McCarthy (1990), who detected (at the $\approx 2\sigma$
level) features at 1300\AA, 1480\AA, and 1840\AA, and
weaker features at 1400\AA, 1640\AA, and 1720\AA. Our spectrum of 3C256
covers the rest wavelength region $\lambda\lambda$1410--3190\AA, but
the only convincing absorption line that we detect is the 1477\AA\ line
described above. A marginal line may be also present at 1712\AA, but
better signal-to-noise ratio spectra are required to confirm this. We
see no evidence of the 1840\AA\ line.  Chambers \& McCarthy interpret
the absorption features that they detect (albeit marginally) in
composite radio galaxy spectra as evidence that the galaxian continuum
of high redshift radio galaxies is dominated by starlight from a young
stellar population.  However, the line that we measure, although
roughly coincident with the feature identified by Chambers and
McCarthy, has an extremely large equivalent width and FWHM compared to most
stellar absorption features. The strongest stellar absorption lines in
O and B stars in this spectral region are the
SiIV$\lambda\lambda$1394,1403 and CIV$\lambda\lambda$1548,1551
resonance doublets, which generally have maximum equivalent widths in
the spectra of {\it individual stars} of $\approx$ 12\AA\ (Panek \&
Savage 1976, Sekiguchi \& Anderson 1987). The identification
of the $\lambda$1477 absorption line with stellar CIV absorption would
imply that the starlight is separated in velocity from the line
emitting gas by $\approx$14300\kms, which appears unrealistic.

The only known stellar absorption lines that are close in wavelength to
our mystery line at 1477\AA\ are weak blends at 1478\AA\ (possible
contributing ions are SrII, SI, SiII, NiII, TiIII, and NI),
1465\AA\ (several blended FeV and FeIV lines), 1457\AA\ (possible
contributing ions are NiII, TiIII, TiIV and SI) and FeV$\lambda$1453
(Underhill, Leckrone, \& West 1972; Kinney \etal 1993; Fanelli \etal
1992, 1987; Dean \& Bruhweiler 1985). All of these features are weak,
and in integrated spectra they rarely have equivalent widths greater
than $\approx 1$\AA\ (Fanelli \etal 1992, Leitherer, Robert, \& Heckman
1995).  In summary, the large equivalent width of the line, the absence
of any known strong stellar feature at $\lambda \approx 1477$\AA, and
the lack of other strong stellar absorption features in the spectrum
imply that the absorption feature is probably not due to a young
population.

The large FWHM of the line might suggest that the feature may arise
from a wind. Broad absorptions in the P-Cygni profiles of CIV and SiIV
lines in some O stars have FWZI as large as 3500\kms (\eg, Cassinelli
1979 and references therein). However, the interpretation of the
1477\AA\ line in 3C256 as CIV absorption due to outflowing material
results in huge outflow velocities ($\approx$14300\kms\ in the line
center!). The possibility that the absorption is a P-Cygni profile
associated with the semi--forbidden NIV]$\lambda$1486 line can also be
ruled out since this requires huge column densities and would therefore
predict very strong P-Cygni profiles associated with the HeII and CIV
lines.

\noindent\underbar{\it Interstellar or Intervening Absorption?}

There are several FeIV and FeV transitions in this spectral region that
have been measured against O star spectra (\eg, Dean and Bruhweiler
1985). It is conceivable that these transitions could contribute to the
absorption, but it is improbable that they would result in absorption
in excess of 12\AA\ in equivalent width, and covering such a large
wavelength range.

One other possibility is that the absorption is due to a foreground,
intervening system, say CIV$\lambda$1549 at z=1.692.  Although there is
no associated MgII$\lambda$2800 absorption at the same redshift, this
in itself is not too unusual since only $\approx$26\% of CIV absorbers
show associated MgII absorption (Steidel \& Sargent 1992).  The more
critical problem with this expanation, however, is that equivalent
width and velocity width 
of the absorption would be unprecedented: most CIV absorption
line systems observed along the lines of sight to quasars have rest
equivalent widths $\simlt$3\AA\ and velocity widths
$\sim$250--300\kms\ (\eg, Steidel \& Sargent 1992).  The only
possibility that this is an intervening absorption system is if the
absorption arises in a forground supercluster similar to that seen
along the line of sight to the quasar pair Tol 1037$-$2703/1038$-$2712
(Jakobsen \etal 1986, Jakobsen \& Perryman 1992). In this case the
broad line would be comprised of several overlapping narrow
absorptions. Even in this scenario, the properties of the foreground
systems would be somewhat extreme: the velocity width of the foreground
supercluster would be nearly 20,000~\kms\ with none of the absorbers
showing associated MgII$\lambda$2800 absorption.

\noindent\underbar{\it BAL Absorption?}

\nopagebreak

The most plausible interpretation of the absorption feature is that it
is broad CIV absorption similar to that observed in BAL QSOs. Unlike in
the preceding discussion where we implicitly assumed that the
absorption arises either in the diluting component or external to the
galaxy, in this scenario the absorption would occur in a 
cloud near the nucleus of the AGN and would therefore be a feature
present in the scattered light spectrum. The large velocity width
(FWHM$\approx$6450\kms) and even the large relative velocity ($\Delta v
\approx 0.047c$) are similar to those seen in BAL QSOs (Korista \etal 1993, 
Turnshek 1988 and references therein). For example, the QSO 0135$-$4001
shows a strong BAL system with a FWHM$\approx$1900\kms\ and a velocity
difference of $\approx$12900\kms\ (Weymann \etal 1981). In addition,
Turnshek (1988) finds that FeII emission may be enhanced and broad CIV
may be weaker in BAL QSOs compared with non--BAL QSOs.  These
correlations may explain the absence (or weakness) of broad CIV and the
presence of moderately strong FeII emission in the polarized flux
spectrum of 3C256. This explanation for the absorption feature is
testable: if the absorption is BAL--like CIV, there should also be
absorption due to SiIV at 3760\AA, NV at 3388\AA, and possibly Lyman
$\alpha$ at 3273\AA.

The one main difficulty with the BAL interpretation for the hidden
nucleus of 3C256 is that the BAL phenomenon is known to strongly
anti--correlate with radio power: all known BAL QSOs are radio
quiet (Stocke \etal 1984, Turnshek 1988). Steep spectrum radio loud
quasars sometimes do show strong absorption features of resonance
lines, but these usually occur at the same systemic velocity as the
emission lines ({\it e.g.,} Foltz \etal 1988). Nevertheless, from a physical 
standpoint, there is no strong reason why BAL clouds should not exist 
in radio galaxies, and there is at least one other known case of 
strong BAL--like absorption associated with a high redshift radio galaxy
(Dey \etal 1996b). 

\subsection {The Stellar Fraction in 3C256}

In this speculative subsection we briefly comment on the possible
stellar content of 3C256.  As already noted in the preceding section,
there is no evidence for any stellar absorption lines in the UV
spectrum of 3C256.  In addition, there is no evidence of the
2600\AA\ and 2900\AA\ breaks that are normally observed in the spectra
of cool stars and old composite populations. If we follow the
definitions of Fanelli \etal (1992,1990) for these spectral breaks
({\it i.e.,} $\Delta_{2600} = -2.5{\rm
log}[F_\lambda(2596-2623)/F_\lambda(2647-2673)]$ and $\Delta_{2900} =
-2.5{\rm log}[F_\lambda(2818-2838)/F_\lambda(2906-2936)]$), then we
find that $\Delta_{2600}\approx 0.01\pm 0.05$ and $\Delta_{2900}\approx
-0.11\pm 0.07$. These values are representative of very early type (O
and B) stars, and therefore imply that either the stellar population is
dominated by hot, young stars, or the fractional contribution of
starlight to the rest frame UV spectrum is very small.

An important implication of the large contribution of scattered AGN
light to the spectrum of 3C256 is that the continuum magnitudes and
colors must be corrected for the AGN light before any conclusions can
be drawn about the underlying stellar population in the galaxy. At
present, the exact fractional contribution of scattered light to the
spectrum is unknown, although in \S~4.1.1 we crudely estimated the
fraction to be 0.1--0.42.  The percentage polarization is observed to be
wavelength independent; if this is due to electron scattering, then 
the scattered light is likely to have the same
spectral shape as the total light spectrum.  Therefore, approximating
the observed optical scattered light spectrum by a powerlaw
$F_\nu\propto\nu^{-1}$ and (under the assumption of electron scattering) 
extrapolating it into the infrared, we find
that the flux due to the scattered AGN component must be $\approx
2-9\mu$Jy at $K$. The observed $K$ flux is $\approx 17~\mu$Jy
($K\approx 19$; Eisenhardt \& Dickinson 1992), and therefore the true $K$--band
($\lambda_{rest}\sim 7800$\AA) magnitude of the underlying stellar
population is 19.1 to 19.8. In comparison, an unevolved $L^*$
(M$_V\approx -21.9$) elliptical at $z=1.8$ would have $K \approx 20.5$;
a typical brightest cluster elliptical like NGC1275 (M$_V=-24.3$;
Oemler 1976) would be at least 2.4 magnitudes brighter.

Therefore, although 3C256 is brighter than an $L^*$ galaxy at the same
redshift, it is significantly (nearly 1.5 magnitudes) fainter than the
mean radio galaxy $K-z$ sequence {\it assuming that the light is
scattered by electrons}. (It is important to note, however, that most
dust scattering models predict that the contribution of scattered light
in the $K$--band is very small. In this case, the galaxy would still be
fainter than the mean $K-z$ relation, but the problem is not as
severe.) 

It is unlikely that 3C256 is an intrinsically low mass galaxy:
the mass estimates derived from the scattered light are comparable to
the masses of present day giant elliptical and cD galaxies. If the main
sequence of the $K-z$ relation defines the locus of evolution of a
massive elliptical and the small scatter observed in the $K-z$ relation
for powerful radio galaxies is indicative that these galaxies have
similar stellar masses, population compositions and histories (Lilly
1989, McCarthy 1993), then the position of 3C256 in the $K-z$ diagram
implies either that 3C256 is much less massive than other powerful
radio galaxies at similar redshifts, or that it has not yet formed the
bulk of its stars. The latter interpretation may also be consistent
with the large baryonic masses required in the interstellar component
to explain the observed scattered luminosity of the galaxy.  We
speculate that if the light in 3C256 is scattered by electrons, then 
3C256 may indeed be a very young galaxy, not because its
luminosity is dominated by a young starburst, but instead because most
of its mass has not yet been converted into stars.

\section{Conclusions}

We have observed polarized spatially extended optical continuum
emission from the high redshift (z=1.824) radio galaxy 3C256. The
percentage polarization is roughly independent of wavelength, and the
position angle of the electric vector is perpendicular to the major
axis of the UV emission, and to the axis defined by the lobes of the
radio source. In addition, the narrow emission lines appear to be
unpolarized and dilute the continuum polarization. There is marginal
evidence that the CII]$\lambda$2326 and the red wing of the
MgII$\lambda$2800 emission lines are slightly polarized, which may
imply some contribution to the lines from a scattered component. These
results support the hypothesis that the radio galaxy contains a buried
quasar that is directed in the plane of the sky and hidden from our
direct view.

We have attempted to determine whether the scattering is dominated by
electrons or by dust. Since electrons are inefficient scatterers,
attributing all of the observed emission to electron scattering results
in a large mass estimate for the ionized component of the interstellar
medium responsible for the scattering. A hot ($T\sim10^7$~K) medium as
suggested by Fabian (1989) can be ruled out based on the observed upper
limit on the soft X-ray flux from 3C256. However, at the present
juncture, we cannot rule out scattering by a cooler population of
electrons; in fact, the electrons required to produce the observed
Ly$\alpha$ luminosity are sufficient to scatter the light if the 
density is low. If electrons are indeed the scatterers, the
dust--to--gas ratio in the scattering region must be much smaller than
(at least $10^{-3}$ times less than) the Galactic value.  If dust
scattering is invoked as the sole explanation of the scattered light,
the dust mass required within the cones is large and, if distributed
evenly, implies that the extinction through the cone would be large and
should be reflected in the ratios of the emission lines. Since the
emission lines do not appear to be highly reddened, the dust scattering
hypothesis requires that the dust and gas is distributed differently:
the dust scatterers are clumped, whereas the line emitting gas is
distributed more evenly in the galaxy.  Hence although dust scattering
is a more efficient process, it may not be any more attractive from an
energetic standpoint than scattering by a population of cool electrons.
In summary, we cannot state conclusively whether the dominant scattering 
population is dust or electrons: both are individually difficult to
reconcile with the observations, at least using a simple filled--cone
geometry. It is very likely, however, that the situations that we have
considered here are overly simplistic, and that in reality both dust
and electrons in a multi-phase medium are jointly responsible for the
scattered light.

We also detect a strong, broad absorption feature located at $\lambda_{rest}
\approx 1477$\AA.  The line cannot be a stellar absorption feature
unless the starlight is separated by $\approx$14300\kms\ in velocity
from the line emitting gas and the stellar spectrum only consists of
strong CIV absorption.  If the absorption line is present in the
scattered flux (and higher signal--to--noise ratio data are required to
check this), then it may arise in the broad line region near the AGN
nucleus.  At present, the identification of this absorption is not
certain, but the most likely explanation is that it is a CIV broad
absorption line system similar to those seen in BAL QSOs. 

Finally, we find no direct evidence for starlight in the observed 
spectrum of 3C256. In fact, if a significant fraction of the observed
continuum emission is scattered AGN light, then the stellar component is 
quite underluminous compared to that observed in other powerful radio 
galaxies at similar redshifts. Since the mass estimates derived from 
the scattering arguments imply that 3C256 is a massive galaxy, we speculate
that it is very young and has not yet formed the bulk of its stars.

\bigskip

We thank Tom Bida and Joel Aycock for invaluable help during our Keck
LRIS run. We are very grateful to Ken Chambers, Sperello di Serego
Alighieri, Peter Eisenhardt, Bob Fosbury, James Graham, Todd Hurt,
Buell Jannuzi, John Kartje, Ari Laor, Joan Najita, and Dave Turnshek for useful
discussions. We thank George Djorgovski for permitting us to use his
images of 3C256 in this paper, and Joan Najita for a careful reading of
the manuscript. We are very grateful to the referee, Dr.\ di Serego
Alighieri, for constructive comments on the manuscript, and for calculating 
dust models to the observed spectrum. We would also like to
thank Joe Miller and Marshall Cohen for their encouragement and support
of this project.  A.\ D.\  is grateful to Paul Ho for his hospitality
at the CfA during the preparation of the manuscript.  R.\ J.\ A.\ 
gratefully acknowledges NSF grant \# AST-9321441.  H.\ S.\ gratefully 
acknowledges NSF grant \# AST-9225133.  The W.\ M.\ Keck Observatory is
a scientific partnership between the University of California and the
California Institute of Technology, made possible by the generous gift
of the W.\ M.\ Keck Foundation. This work was performed at IGPP/LLNL
under the auspices of the U.\ S.\ Dept.\ of Energy under contract \#
W-7405-ENG-48.

\pagebreak

\setcounter{page}{38}

\centerline {\bf References}

\medskip

\refindent Antonucci, R. 1993, \araa, 31, 473.

\refindent Antonucci, R., Hurt, T.\ \& Miller, J.\ 1994, \apj, 430, 210. 

\refindent Antonucci, R.\ J.\ \& Miller, J.\ S. 1985, \apj, 297, 621.

\refindent Appenzeller, I.\ 1968, \apj, 151, 907.

\refindent Barthel, P.\ D.\ 1989, \apj, 336, 606. 

\refindent Becker, R.\ H., White, R.\ L.\ \& Edwards, A.\ L.\ 1991, 75, 1.

\refindent Brotherton, M.\ S., Wills, B.\ J., Steidel, C.\ C., \&
Sargent, W.\ L.\ W. 1994, \apj, 423, 131.
 
\refindent Brown, J.\ C.\ \& McLean, I.\ S.\ 1977, \aa, 57 141.

\refindent Bruzual A., G. 1986, {in Nearly Normal Galaxies}, Proc.\ of the
Eighth Santa Cruz Summer Workshop, ed.\ S.\ M.\ Faber, (Springer-Verlag:
New York) p.\ 265.

\refindent Calzetti, D., Bohlin, R.\ C., Gordon, K.\ D., Witt, A.\ N.\ \& 
Bianchi, L.\ 1995, \apj, 446, L97.

\refindent Cassinelli, J.\ P.\ 1979, \araa, 17, 275.

\refindent Chambers, K.\ C.\ \& McCarthy, P.\ J.\ 1990, \apj, 354, L9

\refindent Chambers, K.\ C., Miley, G.\ K.\  \& Joyce, R.\ R. 1988, \apj, 329, L75.

\refindent Chambers, K.C., Miley, G.K. \& van Breugel, W. 1987, \nature, 329, 604.

\refindent Cimatti, A., Alighieri, S.\ D., Fosbury, R.\ A.\ E., Salvati, M.\
\& Taylor, D.\ 1993, \mn, 264, 421.

\refindent Cimatti, A., Dey, A., van Breugel, W., Antonucci, R.\ \& Spinrad, 
H.\ 1995, submitted to \apj. 

\refindent Crawford, C.\ S.\ \& Fabian, A.\ C.\ 1995, \mn, 273, 827.

\refindent Daly, R.\ A.\ 1992, \apj, 386, L9.

\refindent Dean, C.\ A.\ \& Bruhweiler, F.\ C.\ 1985, \apjsupp, 57, 133.

\refindent Dey, A.\ \& Spinrad, H.\ 1996, \apj, in press.

\refindent Dey, A., van Breugel, W., Antonucci, R.\ \& Spinrad, H.\ 1996a, 
in preparation.

\refindent Dey, A., van Breugel, W., Spinrad, H., R\"ottgering, H., 
Hurt, T., \& Antonucci, R.\ 1996b, in preparation.

\refindent De Young, D.\ S.\ 1981, \nature, 293, 43.

\refindent De Young, D.\ S. 1989, \apj, 342, L59

\refindent Dickinson, M., Dey, A.\ \& Spinrad, H.\ 1995, in proceedings
of Ringberg Conf.\ on ``Galaxies in the Young Universe'',
H.\ Hippelein, ed., in press.

\refindent di Serego Alighieri, S., Binette, L., Courvoisier, T.\ J.--L., 
Fosbury, R.\ A.\ E., \& Tadhunter, C.\ N.\ 1988, \nature, 334, 591.

\refindent di Serego Alighieri, S., Cimatti, A., \& Fosbury, R.\ A.\ E. 1993, \apj, 404, 584.

\refindent di Serego Alighieri, S., Cimatti, A., \& Fosbury, R.\ A.\ E. 1994, \apj, 431, 123.

\refindent di Serego Alighieri, S., Fosbury, R.\ A.\ E., Quinn, P.\ J.\  
\& Tadhunter, C.\ N.\ 1989, \nature, 341, 307.

\refindent Djorgovski, S., Spinrad, H.\ \& Marr, J.\ 1985, in {New
Aspects of Galaxy Photometry}, ed. J.L. Nieto (Springer Verlag: New
York), p.193

\refindent Dunlop, J.\ S.\ \& Peacock, J.\ A.\ 1993, \mn, 263, 936.

\refindent Eales, S.\ A.\ \etal 1993, \apj, 409, 578. 

\refindent Eisenhardt, P. \& Dickinson, M.\ 1992, in The Evolution of 
Galaxies and Their Environment, Proc.\ of the Third Teton Summer School
on Astrophys., NASA CP--3190, p.29.

\refindent Fabian, A.\ C. 1989, \mn, 238, 41P.

\refindent Fanelli, M.\ N., O'Connell, R.\ W., Burstein, D.\ \& Wu, C.-C.\ 1992, \apjsupp, 82, 197.

\refindent Fanelli, M.\ N., O'Connell, R.\ W., Burstein, D.\ \& Wu, C.-C.\ 1990,
\apj, 364, 272.

\refindent Fanelli, M.\ N., O'Connell, R.\ W.\ \& Thuan, T.\ 1987, \apj, 321, 768.

\refindent Foltz, C.\ B., Chaffee, F.\ H., Weymann, R.\ J.\ \&
Anderson, S.\ F.\ 1988, in {QSO Absorption Lines}, STScI
Symp.\ Ser.\ v.2, J.\ C.\ Blades, D.\ Turnshek \& C.\ A.\ Norman, eds.,
(Cambridge Univ.\ Press: Cambridge), p.53.

\refindent Fosbury, R.\ A.\ E., di Serego Alighieri, S., Courvoisier, T., 
Snijders, M.\ A.\ J., Tadhunter, C.\ N., Walsh, J., \& Wilson, W.\ 1990, in 
{Evolution in Astrophysics}, Toulouse, ESA SP-310. 

\refindent Goodrich, R.\ W., Cohen, M.\ H.\ \& Putney, A.\ 1995, \pasp, 107, 179. 

\refindent Gregory, P.\ C.\ \& Condon, J.\ J.\ 1991, \apjsupp, 75, 1011.

\refindent Oke, J.\ B., Cohen, J.\ G., Carr, M., Cromer, J., Dingizian, A., 
Harris, F.\ H., Labrecque, S., Lucino, R., Schaal, W., Epps, H., \& Miller, 
J.\ 1995, \pasp, 107, 375.
 
\refindent Henyey, L.\ G.\ \& Greenstein, J.\ L. 1941, \apj, 93, 70.

\refindent Hummer, D.G. \& Storey, P.J. 1987, \mn, 224, 801.

\refindent Jakobsen, P.\ \& Perryman, M.\ A.\ 1992, \apj, 392, 432.

\refindent Jakobsen, P., Perryman, M.\ A., Ulrich, M.\ H., Macchetto, F.\ 
\& di Serego Alighieri, S.\ 1986, \apj, 303, L27.

\refindent Jannuzi, B.\ T.\ 1994, in Multi Wavelength Continuum Emission of 
AGN, IAU Symp.\ 159, ed.\ T.\ J.--L.\ Courvoisier \& A.\ Blecha, (Kluwer: Dordrecht), p.~470.

\refindent Jannuzi, B.\ T.\ \& Elston, R. 1991, \apj, 366, L69.

\refindent Jannuzi, B.\ T., Elston, R., Schmidt, G., Smith, P., \&
Stockman, H.\ 1995, ApJ, 454, L111.

\refindent Kartje, J. F. 1995, \apj, 452, 565.

\refindent Kinney, A.\ L., Bohlin, R.\ C., Calzetti, D., Panagia, N.\ \& 
Wyse, R.\ F.\ G.\ 1993, \apjsupp, 86, 5.

\refindent Korista, K.\ T., Voit, G.\ M., Morris, S.\ L.\ \& Weymann, R.\ J.\ 
1993, \apjsupp, 88, 357.

\refindent Kriss, G.\ A.\ 1994, in Proceedings of the 3rd Conference on
Astrophysics Data Analysis \& Software Systems, ASP Conf.\ Ser.\ v. 61,
ed.\ D.\ R.\ Crabtree, R.\ J.\ Hanisch, \& J.\ Barnes.

\refindent Laor, A.\ 1995, in preparation.
 
\refindent Le F\`evre, O., Hammer, F., Nottale, L., Mazure, A., \& Christian, C.\ 
1988, \apj, 324, L1.

\refindent Leitherer, C., Robert, C., \& Heckman, T.\ M.\ 1995, \apjsupp, 99, 173.

\refindent Lilly, S.\ J.\ 1989, \apj, 340, 77.

\refindent Lilly, S.\ J.\ \& Longair, M.\ S. 1984, \mn, 211, 833.

\refindent Longair, M.\ S., Best, P.\ N.\ \& R\"ottgering, H.\ J.\ A.\ 1995, 
\mnras, 275, L47.

\refindent Massey, P.\ \& Gronwall, C.\ 1990, \apj, 358, 344.

\refindent Massey, P., Strobel, K., Barnes, J.V. \& Anderson, E. 1988, \apj, 328, 315.

\refindent Mathewson, D.\ S.\ \& Ford, V.\ L.\ 1970, MemRAS, 74, 139. 

\refindent McCarthy, P.J. 1993, \araa, 31, 639

\refindent McCarthy, P.\ J., van Breugel, W.\ \& Kapahi, V.\ K.\ 1991, \apj, 371, 478.

\refindent McCarthy, P.J., van Breugel, W.J.M., Spinrad, H. \& Djorgovski, S.
1987, \apjlett, 321, L29

\refindent Miller, J.\ S.\ \& Antonucci, R.\ J.\ 1983, 271, L7. 

\refindent Miller, J.\ S.\ \& Goodrich, R.\ W.\ 1990, \apj, 355, 456.

\refindent Miller, J.\ S., Goodrich, R.\ W.\ \& Matthews, W.\ G.\ 1991, \apj, 378, 47.

\refindent Miller, J.\ S., Robinson, L.\ B.\ \& Goodrich, R.\ W.\ 1988, in 
{Instrumentation for Ground-Based Optical Astronomy}, ed.\ L.\ B.\ Robinson,
(Springer-Verlag: New York), p.~157.

\refindent Miley, G.\ K., Chambers, K.\ C., van Breugel, W.\ J.\ M.,
\&  Macchetto, F.\ 1992, \apj, 401, L69.

\refindent Oemler, A.\ 1976, \apj, 209, 693.

\refindent Oke, J.\ B. 1974, \apjsupp, 236, 27.

\refindent Osterbrock, D.\ E. 1989, {Astrophysics of Gaseous
Nebulae and Active Galactic Nuclei}, (University Science Books:
California).

\refindent Panek, R.\ J.\ \& Savage, B.\ D.\ 1976, \apj, 206, 167.

\refindent Rees, M.\ J. 1989, \mn, 239, 1P.

\refindent Rybicki, G.\ B.\ \& Lightman, A.\ P.\ 1979, {Radiative
Processes in Astrophysics}, (John Wiley \& Sons: New York).

\refindent Savage, B.\ D.\ \& Mathis, J.\ S.\ 1979, \araa, 17, 73.

\refindent Schmidt, G.\ D., Elston, R.\ \& Lupie, O.\ L.\ 1992, \aj, 1563.

\refindent Schmidt, M. 1963, Nature, 197, 1040.

\refindent Sekiguchi, K.\ \& Anderson, K.\ S.\ 1987, \aj, 94, 129.

\refindent Simmons, J.\ F.\ L.\ \& Stewart, B.\ G.\ 1985, \aa, 142, 100.

\refindent Spinrad, H.\ 1986, \pasp, 98, 269. 

\refindent Spinrad, H.\ \& Djorgovski, S.\ 1984, \apj, 285, L49.

\refindent Spinrad, H., Filippenko, A.\ V., Wyckoff, S., Stocke,
J.\ T., Wagner, R.\ M.\ \& Lawrie, D.\ G.\ 1985, \apj, 299, L7.

\refindent Steidel, C.\ C.\ \& Sargent, W.\ L.\ W. 1991, \apj, 382, 433.

\refindent Steidel, C.\ C.\ \& Sargent, W.\ L.\ W. 1992, \apjsupp, 80, 1.

\refindent Stocke, J.\ T., Foltz, C.\ B., Weymann, R.\ J.\ \& Christiansen, 
W.\ K.\ 1984, \apj, 280, 476.

\refindent Tadhunter C.\ N., Fosbury, R.\ A.\ E., \& di Serego
Alighieri, S. 1988, in Proc. of the Como Conference, BL Lac Objects,
ed. L. Maraschi, T.\ Maccacaro \& M.\ H.\ Ulrich (Berlin:
Springer-Verlag), 79.

\refindent Tran, H.\ D.\ 1995a, \apj, 440, 565.

\refindent Tran, H.\ D.\ 1995b, \apj, 440, 578.

\refindent Tran, H.\ D.\ 1995c, \apj, 440, 597.

\refindent Turnshek, D.\ A.\ 1988, in {QSO Absorption Lines}, STScI
Symp.\ Ser.\ v.2, J.\ C.\ Blades, D.\ Turnshek \& C.\ A.\ Norman, eds., 
(Cambridge Univ.\ Press: Cambridge), p.17.

\refindent Underhill, A.\ B., Leckrone, D.\ S., \& West, D.\ K.\ 1972, \apj, 171, 63.

\refindent Wardle, J.\ F.\ C.\ \& Kronberg, P.\ P.\ 1974, \apj, 194, 249.

\refindent Weymann, R.\ J., Carswell, R.\ F.\ \& Smith, M.\ G.\ 1981, 
\araa, 19, 41.

\refindent Wilson, A.\ N.\ \& Tsvetanov, Z.\ I. 1994, \aj, 107, 1227.

\vspace{4in}


\vfill
\vfill


\end{document}